\documentclass{aa}
\usepackage{natbib}
\usepackage[varg]{txfonts}
\usepackage[export]{adjustbox}
\usepackage{mathtools}
\usepackage{mathrsfs}
\usepackage{fancyvrb}
\usepackage{breqn}
\usepackage{makecell}
\usepackage{pdflscape}
\usepackage{geometry}
\usepackage[normalem]{ulem}

\usepackage{color}

\begin{document}

\title{Unraveling the Formation History of the Black Hole X-ray Binary LMC X-3 from ZAMS to Present}

\author{Mads S\o rensen\inst{1}
  \and Tassos Fragos\inst{1}
  \and James F. Steiner\inst{2}
  \and Vallia Antoniou\inst{3}
  \and Georges Meynet\inst{1}
  \and Fani Dosopoulou\inst{4}
  }
\institute{Observatoir de Genéve, University of Geneva, Route de Sauverny, 1290 Versoix, Switzerland 
\and MIT Kavli Institute for Astrophysics and Space Research, Cambridge, MA 02139, USA
\and Harvard-Smithsonian Center for Astrophysics, 60 Garden Street, Cambridge, MA 02138, USA
\and Center for Interdisciplinary Exploration and Research in Astrophysics (CIERA) and Department of Physics and Astrophysics, Northwestern University, Evanston, IL 60208}

\offprints{M. S\o rensen, \email{mads.sorensen@unige.ch}}

\abstract {} {We have endeavoured to understand the formation and evolution of the black hole (BH) X-ray binary LMC X-3. We estimate the properties of the system at 4 evolutionary stages: 1) at the Zero Age Main Sequence (ZAMS), 2) just prior to the supernova (SN) explosion of the primary, 3) just after the SN, and 4) at the moment of RLO onset.} {We use a hybrid approach, combining detailed stellar structure and binary evolution calculations with approximate population synthesis models. This allows us to estimate potential natal kicks and the evolution of the BH spin. In the whole analysis we incorporate as model constraints the most up-to-date observational information, encompassing the binary's orbital properties, the companion star mass, effective temperature, surface gravity and radius, as well as the black hole's mass and spin.} {We find that LMC X-3 began as a ZAMS system with the mass of the primary star in the range $M_{\mathrm{1,ZAMS}}$ = 22-31 $\mathrm{M_{\odot}}$ and a secondary star of $M_{\mathrm{2,ZAMS}} = 5.0-8.3M_{\odot}$, in a wide ($P_{ZAMS} \gtrsim 2.000\, \rm days$) and  eccentric ($e_{\mathrm{ZAMS}} \gtrsim 0.23$) orbit. Just prior to the SN, the primary has a mass of $M_{\mathrm{1,preSN}} = 11.1-18.0\,\rm  M_{\odot}$, with the secondary star largely unaffected. The orbital period decreases to $0.6-1.7\, \rm days$, and is still eccentric $0 \leq e_{\mathrm{preSN}} \leq 0.44$. We find that a symmetric SN explosion with no or small natal kicks (a few tens of $\rm km\, s^{-1}$) imparted on the BH cannot be formally excluded, however, large natal kicks in excess of $\gtrsim 120 \,\rm km\, s^{-1}$ increase the estimated formation rate by an order of magnitude. Following the SN, the system has a BH $M_{\mathrm{BH,postSN}} = 6.4-8.2\,\rm M_{\odot}$ and is put into an eccentric orbit. At the RLO onset the orbit is circularised and it has an orbital period of $P_{\mathrm{RLO}} = 0.8-1.4\,\rm days$.} {}
\maketitle
%*** Introduction

%Start with a few of paragraphs that are at low level, wikipedia style, where you introduce: 
%1) what X-ray binaries are, 2) what is the standard formation channel that we consider, 
%3) why is it useful to try to reconstruct the evolutionary history of BH XRBs, 
%4) summarize the work that has being done so far (earlier papers from the same project, my LMXB spin paper, the recent paper on M82 X-2 (http://adsabs.harvard.edu/abs/2015arXiv150102679F) Repetto, S., Davies, M. B., & Sigurdsson, S. (2012)., other similar work)
%5) Try to make a point of why LMC X-3 is especially interesting: different metallicity, good distant measurement which results precise and robust measurements. Rare class of intermediate-mass X-ray binaries. Is it transient or persistent?

%6) Mention very briefly in one paragraph what is the goal and methodology of the paper.
%7) End with a paragraph where you mention what is in each of the next sections.
\section{Introduction}
%1)

X-ray binaries (XRBs) are evolved stellar binary systems containing a compact object (CO), a neutron star (NS) or a black hole (BH), and a companion star that is losing mass. The companion star, also known as the donor star, loses mass either because it is overflowing its Roche lobe or through stellar winds. A fraction of the lost mass is transferred to the CO that upon accretion emits X-rays, hence their name. In consequence of the mass transfer (MT) the systems orbital period changes.

Roche-lobe overflow (RLO) is initiated either when the donor star expands due to its stellar interior being restructured, e.g. the star leaves the Main Sequence (MS) and expands outside its Roche lobe, or because angular momentum loss processes (e.g. gravitational wave radiation and magnetic braking) shrink the orbit. RLO triggers MT, through the first Lagrangian point, directly into the potential well of the CO. Stellar winds on the other hand play a less dominant role in shrinking the orbit and are relevant mostly for XRBs with massive donors which are able to generate winds sufficiently powerful to cause a high mass loss, such as those observed in Wolf Rayet stars or red giants \citep[e.g.][]{VKL2001,VK2005, vanLoon:2005ik}. For hot stars, due to stellar winds being nearly spherically symmetric and fast, of the order of $\sim$1000 km\,s$^{\mathrm{-1}}$, Bondi-Hoyle accretion \citep{1952MNRAS.112..195B} of these stellar winds onto the CO is far less effective as a MT mechanism to power an XRB relative to RLO. However, fast wind Bondi-Hoyle accretion affects not only the orbital separation of the binary but may also enhance the system's eccentricity for a mass-ratio < 0.78 \citep[][]{dosopoulou2016a, dosopoulou2016b}.

%3) why is it useful to try to reconstruct the evolutionary history of BH XRBs
Since their first discovery, BH XRBs have proven an essential natural laboratory to study stellar BHs. These systems provide the best observational test for the existence of BHs formed from stellar collapse and the only type of systems where the properties of a stellar BH, i.e. its mass and angular momentum, can be measured based on observation in the electromagnetic spectrum \citep{cowley1992, remillard2006, mcclintock2013}. Furthermore, they provide us with unique observational constraints for binary evolution and CO formation models. 

%4) previous studies on XRB and LMC X-3
The growing sample of Galactic and extragalactic BH XRBs instigated a number of studies about their formation. However, most of them either focused on the details of individual evolutionary phases or made comparisons of evolutionary models with only a subset of the available observational data \citep[e.g.][and references therein]{PRP2002,PRH2003,2006MNRAS.366.1415J,Ivanova2006b,Fragos2010,Li:2015ic, 2015arXiv151002093N}.

Given our knowledge of the currently observed properties of a XRB system, it is in principle possible to reconstruct its evolution from the moment the system was comprised of two ZAMS stars, through the intermediate phases: primary to secondary MT, the common envelop (CE) phase, mass loss and kicks during the SN, including changes in the systems orbital characteristics, and ending with the state of the system at the initiation of its current MT phase that causes its X-ray emission. %As the system moves away from the ZAMS, it enters the first MT phase, then the common envelope (CE) phase, and the collapse of the primary star that formed a CO including potential changes in the systems orbital characteristic due to mass loss and kicks during SN, after which the XRB phase is eventually reached.

For the analysis of the Galactic low-mass XRB (LMXB) GRO J1655-40, \citet{willems2005} developed for the first time a methodology which combined all the aforementioned aspects, including the modeling of the MT phase, the secular evolution of binary prior to RLO, the CO formation and core-collapse dynamics, and the motion of the system in the Galactic potential, into one comprehensive analysis. The results of this analysis were compared to \textit{all} available observational constraints, including the three-dimensional position and velocity of the system in the Galaxy. Following a methodology similar to \citet{willems2005}, several systems have since been systematically analysed \citep{Fragos2009, valsecchi2010, wong2012, Wong2014}.

One aspect of CO formation that has been recognized as an important element, but is not yet understood from first physical principles concerns the effect of asymmetries associated with the core-collapse process and the resulting natal kicks imparted to NSs and BHs giving them very high peculiar velocities, i.e. the center of mass velocity relative to the local standard of rest. For the case of NSs, kinematic observations of radio pulsar populations strongly suggest that NSs acquire very significant natal kicks of the order of a few to several hundreds of km\,s$\mathrm{^{-1}}$ \citep[e.g.][]{Hobbs:2005be}. More recently, evidence from studies of binary systems of two NSs \citep[e.g.][]{2005PhRvL..94e1102P,2006PhRvD..74d3003W} indicates that there may be a subset of NSs acquiring significantly smaller kicks, of the order of tens of $\rm km\,s^{-1}$. A proposed explanations for these smaller kicks is that these NSs formed through an electron capture SN instead of an Fe core-collapse \citep{2004ApJ...612.1044P,2005MNRAS.361.1243P, Linden:2009gx}.
Similarly, the Be/X-ray binaries (i.e. HMXBs with NS compact objects and Be-type star companions) in both Magellanic Clouds show transverse velocities of up to few tens of km/s (<~15-20 km/s; \cite{Antoniou:2010jy, Antoniou:2016iw} with LMC HMXBs traveling with up to ~4 times larger velocities than their SMC counterparts.

Our understanding of BH formation has evolved significantly in the recent years, however, compared to NS, observational evidence of natal kicks is scarce \citep{2001Natur.413..139M,2002A&A...395..595M,2014ApJ...796....2R}. Natal kicks are important as they are an indicator for the formation process of the BH. A large inferred natal kick ($\gtrsim 100\,\rm km\,s^{-1}$) is evident of a core collapse process accompanied by a supernova explosion with asymmetric ejecta, while natal kicks of $\sim 10\,\rm km\,s^{-1}$ indicate a symmetric supernova explosion with perhaps some asymmetries in the neutrino emission, or even BH formation via direct collapse. The BH in the Galactic LMXB XTE J1118+480 is currently the only BH for which an asymmetric natal kick, in the range 80--310\,km~s$^{\mathrm{-1}}$, has been inferred to not only be plausible but required to explain the formation of the system \citep{Fragos2009}. \citet{willems2005} found that the BH in GRO J1655-40 likely formed with a kick of up to 210\,km~s$^{\mathrm{-1}}$, which led to a kick of the system's center of mass of 45--115\,km~s$^{\mathrm{-1}}$. However, a zero kick provides also a solution marginally consistent with the observed properties of the system. The system V404 Cyg also show a high peculiar velocity which suggest 47--102 \,km~s$^{\mathrm{-1}}$ and indicate that this system also received a kick at BH formation, either from mass loss or during the SN \citep{MillerJones:2009ke, MillerJones:2009cs}. \citet{Repetto2012} studied the Galactic population of LMXBs, using analytical estimates of their evolutionary history and following the orbits of the systems in a Galactic potential, which showed that their large vertical spread relative to the Galactic plane could be explained by introducing a natal kick at the BH formation, and derived constraints on the possible kick imparted on the BH of each system in general agreement with previous studies of individual systems. Recently, \citet{2015MNRAS.453.3341R} followed up on their first study and argued the 7 LMXB system they considered received a natal asymmetric kick; five of these sources with relative small kicks and two with kicks of several hundreds km~s$^{\mathrm{-1}}$. However, \citet{2015arXiv151003871M} argued that this claim by \citet{2015MNRAS.453.3341R} is based on a problematic assumption of the likely direction of the kick and the Galactic dynamics of any LMXB born in the thin disc, suggesting that natal asymmetric kicks in excess of $\sim$100\,km~s$^{\mathrm{-1}}$ is unlikely.

Studies of the evolutionary history of high-mass XRBs (HMXBs) have also been carried out. \citet{valsecchi2010} studied the formation of the massive BH XRB M33 X-7 and found that an asymmetric kick was possibly imparted onto the BH during the core-collapse. However, such a kick is not required in order to explain the formation of the system. \citet{Wong2014} reached similar conclusions for IC10 X-1, concluding that the BH in this system cannot have received a kick larger than 130\,km~s$^{\mathrm{-1}}$, while the analysis of the evolutionary history of Cyg X-1 resulted in an upper limit for a potential asymmetric kick onto the BH  of <77\,km\, s$^{\mathrm{-1}}$ \citep{wong2012}.

Over the last decade, measuring the spins of stellar-mass BHs became possible, by three different methods: the thermal X-ray continuum fitting method, the iron $K_{\alpha}$ method, and the analysis of the quasi-periodic oscillations (QPOs) of the X-ray emission. For an extensive review of the different methods and the so far available measurements see \citet{mcclintock2013}, \citet{Reynolds2013}, \citet{Motta2014b} and references therein. The measured values of the spin parameter $a_{\mathrm{\ast}}$ for the different BHs cover the whole parameter space from $a_{\mathrm{\ast}}$ = 0 (non-rotating BHs) all the way to $a_{\mathrm{\ast}}$ = 1 (maximally rotating BHs), where $a_{\mathrm{\ast}} \equiv cJ/GM^2$ with $|a_{\mathrm{\ast}}| \le 1$. Here $M$ and $J$ are the BH mass and angular momentum respectively, $c$ is the speed of light in vacuum and $G$ the gravitational constant. The question that naturally arises is what is the origin of the measured BH spin. \citet{2015ApJ...800...17F} showed that the spin of known Galactic LMXBs can be explained by the mass accreted by the BH after its formation. This also implies that the natal BH masses in these systems are significantly different from the currently observed ones. In contrast BHs in HMXBs have spins which cannot be explained through accretion after the BH formation due to their short lifetimes and relatively low mass accretion rates. Hence their spin is likely natal \citep{valsecchi2010,Gou2011}. However, see also \citet{Mendez2008, Mendez2011a, Mendez2011b} who suggest that BH HMXB can obtain their spin after formation, through a short period of extreme mass accretion, so-called hypercritical accretion, orders of magnitudes above the critical Eddington accretion rate.

%Besides investigating BH spin and its origin another important aspect to study in relation to XRBs is the combined effect of mass loss and kicks imparted during CO formation as these affects the binary systems orbital parameters. Mass loss results in loss of binding energy which can be countered by an out of phase kick from an asymmetric explosion, where by the binary system survives the loss of mass. In consequence, the binary system orbital parameters, both in the corotating frame and the systemic center of mass motion is affeced\citep{kalogera1996, willems2005}. 

%5)Why LMC X-3 is interesting
LMC X-3 is an intriguing accreting BH binary that stands out compared to the rest of the Roche-lobe overfilling, dynamically confirmed, BH XRBs. The donor star in LMC X-3 is a thermally disturbed, early B type star, making it the most massive among known Roche-lobe overfilling BH XRBs \citep{orosz2014}. This classifies it in the elusive group of intermediate-mass XRBs (IMXBs). As the donor stars of IMXBs are losing mass, they quickly evolve to the significantly longer lived phase of LMXBs. This naturally explains their overall rarity. The fact that we observe such a system in a small galaxy like the Large Magellanic Cloud (LMC), can be explained by the recent star-formation history of the LMC, see Sec. \ref{sec:IMXB}. %which peaks close to the inferred age of LMC X-3. {\blue Valia, can you please validity this sentence, and if possible add some numbers and references?}. 
Being a member of the LMC also offers the advantage of a very well determined distance. LMC X-3 has always been bright, albeit highly variable, in X-rays since the first X-ray telescope observed LMC back in the 70s \citep{leong1971}. This unusual behavior in its X-ray emission again puts LMC X-3 apart from other ``typical'' BH LMXBs \citep{steiner2014x-ray_lag}. Therefore, it is important to place LMC X-3 within the frame of its relatives to see how this system has formed and evolved as well as trying to understand its current nature of being highly variable.

%7) The structure of our paper
In the present paper we investigate the past evolution of LMC X-3 from a ZAMS binary, through intermediate phases and until the current XRB phase. We follow the evolution of LMC X-3 using a hybrid approach, where we combine detailed stellar structure and binary evolution calculation with more approximate population synthesis calculation, in an approach similar to \citet{2015ApJ...802L...5F}. Using the stellar evolution code Modules for Experiments in Stellar Astrophysics \citep[MESA]{paxton2011,paxton2013,2015ApJS..220...15P}, we first scan a 4-dimensional parameter space for potential progenitor systems (PPS) of LMC X-3 at the onset of RLO. We then continue with the parametric binary evolution code BSE \citep{Hurley2000,hurley2002}, to conduct a population synthesis study starting at the ZAMS and evolve a large set of binaries until the onset of RLO, matching the outcome of BSE with that of MESA in order to find the PPS of LMC X-3 at ZAMS. In doing so, we also estimate potential natal kicks and the evolution of BH spin. In the whole analysis we incorporate as model constraints the most up-to-date observational information of the system.

The layout of the rest of the paper is as follows. In Sect. 2 we consider the observational history and currently available observational constraints on the properties of LMC X-3. Section 3 describes how we model the current MT phase from RLO onset to today and search a grid of MT sequence models to fit the observational constraints. We end Sect. 3 by presenting the result of MT sequences describing the LMC X-3 at the onset of RLO. In Sect. 4 we investigate the past evolution, from ZAMS binary system until the initiation of the second MT onto the BH with a population synthesis study which we combine with the results of Sect. 3. In Sect. 5 we discuss our results and in section 6 draw our conclusions.
%*** Observational Constrains
%
%1) You can start with a historical review of of LMC X-3 discovery and first observations
%2) A table with all the observational constraints we are going to use. In the text that refers to the table mention explicitly that we are considering 2-sigma errors.
%3) You can also mention any other interesting observational properties.  
%
%
% Any characterisation of LMC X-3 uses the mass function from which it derives the components of the system.
%
%LMC X-3 variable in visual band due.
%

\section{Observational Constraints}\label{sec:obs}
Around New Year's Day of 1971, the X-ray satellite UHURU looked in the direction of the LMC and found 3 point like sources which were designated LMC X-1, LMC X-2, and LMC X-3\citep{leong1971}. With the next generation X-ray satellite Copernicus, the existence of the three point sources within the LMC was confirmed and their position were further constrained. Furthermore, it was suggested that the sources are binary systems with optical counterparts orbiting a CO. Three candidates were suggested as the optical companion for LMC X-3 \citep{rapley1974}. In a search for an optical companion of LMC X-3, \citet{warren1975} located several stars within the error circle in the direction of LMC X-3. Using colour-colour diagrams these were condensed down to one candidate star of luminosity class III-IV.

Using spectroscopic observations, \citet{cowley1983} suggested the companion was a class B3V, and found a radial velocity confirming it as the optical counterpart of LMC X-3 with an orbital period $\sim$1.70 days. \citet{cowley1983} was also able to estimate the binary mass function to be 2.3 $M_{\mathrm{\odot}}$ with a donor mass $M_{2}$ = 4--8 $\mathrm{M_{\odot}}$ and CO mass $M_{1}$ = 6--9 $\mathrm{M_{\odot}}$, which indicates that the CO is a BH, the first extra galactic stellar-mass BH of its kind. It has also been claimed that the CO in LMC X-3 was an over-massive NS \citep{mazeh1986}, however, the BH model remained the most convincing model to describe the system. Reviewing literature on LMC X-3, \citet{cowley1992} argued its CO to be a BH with a minimum mass of 5 $\mathrm{M_{\odot}}$.

\citet{vanderklis1983} also found a period of $\sim$1.70 days due to ellipsoidal variations in the observed light curve. Due to the first classification of the donor as a giant star, the BH accretion disc was interpreted to be driven by stellar winds rather than RLO. The donor luminosity classification was revised by \citet{soria2001} who found the donor star to more likely be a B5 sub-giant hereby suggesting that the system sustains its accretion disc through RLO.

Recently, \citet{orosz2014} found the classification of the donor star to be difficult to assess as its mass is below that of a single B5 star, but its surface gravity fits such a description, leaving the spectral classification of the donor star as an open question. Hence, the assumption of using isolated star spectral models in order to describe binary star members in a mass-transfer system potentially overestimates the mass of the star. As to the mechanism driving the accretion disc, \citet{orosz2014} also find LMC X-3 to be a RLO mass-transferring system, though its donor star is atypically massive compared to other RLO XRBs, making the system an RLO IMXB.

%Next step is to include articles on accretion disk, main subject, variations due to warping precession or due to mass transfer process.
The long time baseline of LMC X-3 observations, in both the X-ray and optical/infrared bands, shows the system to vary in luminosity by as much as three orders of magnitude, and there have been suggestions for a super-orbital X-ray periodicity of 99--500 days which is caused by either precession of a warped accretion disc \citep{cowley1991} or due to variability of the BH accretion rate \citep{brocksopp2001}. More recent analysis and modelling of the LMC X-3 accretion disk do not find a super-orbital periodicity related to precession of a warping disk or other forms of orbital dynamics \citep{steiner2014a} and most probably the observed variability is due to variations in the accretion rate.

%Finally end up with the latest determination of parameters. Don't forget the theoretical papers.
The most recent set of  parameters for LMC X-3 is given in Table~\ref{tab:LMC_X-3} and is based on the currently most up-to-date data available \citep{ orosz2014, steiner2014}. These new estimates find that the system has an orbital period $P_{orb}$ = 1.7048089 days, a donor star of mass $M_{\mathrm{2}}$ = 3.63 $\pm$ 0.57~$\mathrm{M_{\odot}}$ and effective temperature $T_{eff}$ = 15250 $\pm$ 250 K, and a BH of mass $M_{\mathrm{BH}}$ = 6.98 $\pm$ 0.56$\mathrm{M_{\odot}}$ and spin $a_{\ast}$ = 0.25 $\pm$ 0.12. Compared to earlier estimates, the potential range of masses for both $M_{\mathrm{2}}$ and $M_{\mathrm{BH}}$ are significantly reduced. The distance D = 49.97 $\pm$ 1.3 kpc to LMC X-3 is here taken from the distance to LMC \citep{pietrzynski2013}.

\begin{table*}[!htb]
\centering
\caption{Adopted observed properties of LMC X-3. }\label{tab:LMC_X-3}
\begin{tabular}{clllc}
\hline \hline
j\tablefootmark{a} & Parameter & Notation & Value, ($\mu_{\mathrm{j}} \pm \sigma_{\mathrm{j}}$) & References						\\ 
\hline
& Right Ascension 					  		& RA (h:m:s)					& 05:38:56.63	$\pm$  0.05"    & 3 \\
& Declination 							    & De (d:m:s) 							    & -64:05:03.29	$\pm$	0.08"			& 3 \\
& Distance    		        	            & D (kpc)			  							& 49.97 $\pm$ 1.30	  	    & 4 \\
& & & & \\
& Orbital period             			  	& $P_{\mathrm{orb}}$ (days) 	  					        & 1.7048089 					& 1 \\
  & Inclination                  			  	& i ($\mathrm{^{\circ}}$)        	  			&  69.24 $\pm$ 0.727   & 1 \\
  & Orbital separation      			  		& $a$ ($\mathrm{R_{\odot}}$)                  	& 13.13 $\pm$ 0.45   	& 1 \\
1 & BH mass                    			  	& $M_{\mathrm{BH}}$ ($\mathrm{M_{\odot}}$)    	&   6.98 $\pm$ 0.56  		& 1 \\
2 & Mass ratio $M_{\mathrm{1}}/M_{\mathrm{2}}$	& $q^{\mathrm{-1}}$                 		&   1.93  $\pm$ 0.20    	& 1 \\
  & Donor mass               			  	& $M_{\mathrm{2}}$ ($\mathrm{M_{\odot}}$)      	&   3.63 $\pm$ 0.57  		& 1 \\
3 & Donor Radius              			  	& $R_{\mathrm{2}}$ ($\mathrm{R_{\odot}}$)         & 4.25 $\pm$ 0.24    		& 1 \\
4 & Donor surface gravity 			  	    & $\log g_{\mathrm{2}}$ (cgs) 	  				& 3.740 $\pm$ 0.020    & 1 \\
5 & Donor effective temperature 	        & $T_{\mathrm{eff}}$ (k)		  			    & 15250 $\pm$ 250 k 	& 1 \\
  & BH spin 								& $a_{\ast}$ 	  								& 0.25 $\pm$ 0.12 		& 2 \\
\hline
\end{tabular}
\tablefoot{ \\
\tablefoottext{a}{j is the index of the parameters that are used in estimating the likelihood of a calculated mass transfer being the progenitor of LMC X-3 given its current observed properties. See Sect. \ref{sec:MTsequence}} for details.}
\tablebib{(1)~\citet{orosz2014}; (2) \citet{steiner2014}; (3) \citet{Cui:2002cm}; (4) \citet{pietrzynski2013}}

\end{table*}

%*** MT calculations
%Here you could have 4 subsections:
%
%-- The description of the code and the setup you use, along the choice of parameters. 
%
%-- A description of the finite part of the parameter space that we have to cover. The initial scan of the parameter space with the fiducial MT calculation.
%
%-- Describe how you search for winners, and show a few different indicative MT sequences describing qualitative how the MT sequences behave in different part of the parameter space.
%
%-- Present a summary of all the winning MT sequences and have one summary table of the derived constraints at the onset of RLO, and a full table with all the winning sequences one by one. Only a part of the table will be in the actual table and the full %version will be only online

\section{Modelling the X-ray Binary Phase}\label{sec:MTsequence}
In this first part of the analysis we scan a 4-dimensional parameter space to find the PPS of LMC X-3 at RLO onset. Our free parameters are the BH mass $M_{\mathrm{BH,RLO}}$, the mass ratio $q$, and the orbital period $P_{\mathrm{RLO}}$ at the onset of RLO, as well as the accretion efficiency parameter $\beta$. %that denotes the fraction of the mass funneled towards the accretor that is leaving the system in the form of fast disk winds or jets, carrying the specific orbital angular momentum of the accretor. 
Because of the significant computational cost of a detailed MT calculation, it is practically impossible to fully cover the 4-dimensional parameter space without first limiting the range of parameter values that we need to explore. In order to do so, for every combination of $M_{\mathrm{BH,RLO}}$,  $M_{\mathrm{2,RLO}}$,  $P_{\mathrm{RLO}}$ and $\beta$ we consider, we first use an analytic point-mass MT model to follow the evolution of the orbit. If a set of initial conditions passes the first analytic test, we then perform a detailed calculation of the MT sequence using the stellar evolution code MESA. Finally, we search each detailed simulation for the physical characteristics observationally inferred for LMC X-3.

\subsection{Analytic Point-Mass Mass-Transfer Model}\label{sec:analytic_model}

%In order to limit our parameter space, at least with respect to the secular evolution of the binary system, we use a computationally cheap analytical model point-mass MT model to exclude the largest part of the initial parameter space. Only the remaining volume of the parameter space is then scanned using detailed stellar structure and binary evolution calculations. This allows us to perform a more concentrated search in a well defined region and gives a better resolution to the relevant part of the grid search.

Our analytic point-mass MT model follows the secular evolution, due to RLO MT, of a binary system composed of two point masses. For the simple model, we ignore the effects of stellar interior structure and evolution of the donor, and only focus on mass being transferred and lost in the system, as well as the evolution of the orbit and the BH spin.

We assume that the two point masses, $M_{\mathrm{BH}}$ and $M_{\mathrm{2}}$, are in a circular orbit around their mutual center of mass. In a co-rotating frame with origin at the center of mass, $r_{\mathrm{BH}}$ and $r_{\mathrm{2}}$ are the distances of point masses $M_{\mathrm{BH}}$ and $M_{\mathrm{2}}$ from their mutual center of mass, orbiting each other at a separation $a = r_{\mathrm{BH}} + r_{\mathrm{2}}$. Then $M_{\mathrm{BH}}r_{\mathrm{BH}} + M_{\mathrm{2}}r_{\mathrm{2}} = a(M_{\mathrm{BH}} + M_{\mathrm{2}})$ relates the point masses with their orbital separation.

The angular momentum around the center of mass of the system with eccentricity $e$, and angular velocity $\omega$ is
\begin{equation}
J = J_{BH} + J_{2} = (M_{BH} r_{BH}^2 + M_{2} r_{2}^2 ) \omega\sqrt{1 - e^2} = \mu a^2\omega \sqrt{1-e^2},
\label{eq:angular_momentum}
\end{equation}
where $\mu = \tfrac{M_{\mathrm{BH}}M_{\mathrm{2}}}{M_{\mathrm{BH}} + M_{\mathrm{2}}}$ is the reduced mass. The orbital period, $P$, is related to the systems orbital separation through Kepler's Third Law by 
\begin{equation}\label{eq:orbital_period}
P = \frac{2\pi}{\omega} = 2\pi\sqrt{\frac{a^3}{G(M_{BH} + M_{2})}}
\end{equation}
where
\begin{equation}
\omega = \sqrt{\frac{G(M_{BH} + M_{2})}{a^3}}
\end{equation}
The effect of a change in angular momentum on the orbit separation is given as
\begin{equation}
\frac{\dot{a}}{a} = 2 \frac{\dot{J}}{J} - 2\frac{\dot{M_{BH}}}{M_{BH}} - 2\frac{\dot{M_{2}}}{M_{2}} + \frac{\dot{M_{BH}} + \dot{M_{2}}}{M_{BH} + M_{2}}
\label{eq:angular_momentum_change}
\end{equation}
where a single dot defines the first order derivative with respect to time.

In the analytic point-mass MT model, exchange and loss of angular momentum happens due to exchange of mass between the two point masses, or by the removal of mass from one of the two point masses. When mass is transferred to or lost from the vicinity of a point mass, it carries the specific orbital angular momentum $j = r\omega^{\mathrm{2}}$ of that point mass. We only consider RLO through the Lagrangian point L1, which in general is the dominating MT mechanism in LMXBs and IMXBs \citep{tauris2006}.
During RLO, a fraction $\alpha$ of the mass lost from the donor ($\dot{M}_{\mathrm{2}}$) will escape the system en route to the accretor carrying the specific orbital angular momentum of the donor; i.e., angular momentum exchange in the accretion flow is ignored. The remaining fraction 1-$\alpha$ is funnelled through the L1 point towards the accretor. A fraction $\beta$ of the mass transferred through the L1 point, i.e. $\beta(1-\alpha)\dot{M}_{\mathrm{2}}$, will be lost from the system from the vicinity of the accretor, carrying its specific angular momentum. The remaining fraction $(1-\beta)(1-\alpha)\dot{M}_{\mathrm{2}}$ will be accreted onto the accretor. The mass change of the BH is $\dot{M}_{\mathrm{BH}} = -(1-\alpha)(1-\beta)\dot{M}_{\mathrm{2}}$ and the time derivative of the angular momentum is given as
\begin{equation}
\dot{J} = \alpha\dot{M_{2}}r_{2}^2\omega + (1-\alpha)\beta\dot{M_{2}}r_{BH}^2\omega
\end{equation}

As we only consider transfer of mass we can ignore the time dependent mass change rate $\dot{M}_{\mathrm{2}}$ and instead use the transformation $\tfrac{d}{dt} = \tfrac{dM_{\mathrm{2}}}{dt}\frac{d}{dM_{\mathrm{2}}}$ to get the orbital evolution as a function of $M_{\mathrm{2}}$.
The orbital evolution given by eq. \eqref{eq:angular_momentum_change} of the system during RLO MT ignoring  other potential angular momentum loss mechanisms, then becomes:
\begin{equation}
\frac{da}{dM_{2}} = \frac{2a}{M_{2}}\left[\alpha(1-q)-(1-q)+\frac{q}{2(1+q)}(\alpha \beta- \alpha - \beta)\right].
\label{eq:dadM2}
\end{equation}
We solve this equation analytically to get the orbital period $P$ as a function of the donor mass $M_{\mathrm{2}}$ and use eq. \eqref{eq:orbital_period} to get:

\begin{equation}
\begin{split}
& \frac{P(M_{2})}{P_{\mathrm{RLO}}} = \\
& \left(\frac{M_{2}}{M_{2,RLO}}\right)^{3(\alpha-1)}\left(\frac{M_{\mathrm{BH}}}{M_{BH,RLO}}\right)^{\tfrac{3}{\beta-1}} \left(\frac{M_{2,RLO}+M_{BH,RLO}}{M_{2}+M_{BH}}\right)^{2}
\end{split}
\label{eq:orbit_evo}
\end{equation}

where $M_{\mathrm{BH}} = M_{\mathrm{BH,RLO}} + (1-\beta)(1-\alpha) (M_{\mathrm{2,RLO}}-M_{\mathrm{2}})$ and the subscript RLO refers to initial system values at the onset of RLO. Eq. \eqref{eq:orbit_evo} is only valid for $\beta \neq 1$, $M_{\mathrm{2,RLO}} \neq 0$ and $M_{\mathrm{BH,RLO}} \neq 0$. For $\beta = 1$ eq. \eqref{eq:dadM2} is reduced and a new analytic solution can be found. For the case where $\beta = 1$ the time dependent mass change rate of the accreting BH vanishes and the solution to eq. \eqref{eq:dadM2} then is
\begin{equation}
\begin{split}
& \frac{P(M_{2})}{P_{RLO}} = \\
&\left(\frac{M_{2}}{M_{2,RLO}}\right)^{3(\alpha-1)}e^{\left(\frac{3(1-\alpha)}{M_{BH}}(M_{2}-M_{2,RLO})\right)}\left(\frac{M_{2,RLO}+M_{BH,RLO}}{M_{2}+M_{BH}}\right)^{2}
\end{split}
\label{eq:orbit_evo_no_beta}
\end{equation}

For each combination of initial parameters $P_{\mathrm{RLO}}$, $M_{\mathrm{2,RLO}}$, $M_{\mathrm{BH,RLO}}$, $\alpha$ and $\beta$, we are finding the root of the equation $P(M_{\mathrm{2}})=P_{\mathrm{obs}}$, where $P_{\mathrm{obs}}$ is the currently observed orbital period of LMC X-3. The root of this equation, if it exists, gives us the mass $M_{\mathrm{2,cur}}$ that the donor has when the orbit reaches the observed period. From this, one can also calculate the current mass of the BH as $M_{\mathrm{BH,cur}} = M_{\mathrm{BH,RLO}} + (1-\beta)(1-\alpha) (M_{\mathrm{2,RLO}}-M_{\mathrm{2,cur}})$. From the current mass of the BH and the donor we get the mass ratio $q$. If the calculated $q$ and $M_{\mathrm{BH,cur}}$ are further away than two standard deviations from the observed current values (see Table \ref{tab:LMC_X-3}) then this combination of $P_{\mathrm{RLO}}$, $M_{\mathrm{2,RLO}}$, $M_{\mathrm{BH,RLO}}$, $\alpha$ and $\beta$ values is excluded from a further search of the parameter space, as a binary with these properties at the onset of RLO is not a PPS of LMC X-3.

%Possible solutions of eq. \eqref{eq:orbit_evo} are presented as white areas in figure \ref{fig:MTseq_no_Edd} and is discussed later in this section. The analytical point-mass MT model is an effective method to reduce the potential solution space for the considered parameters. With the additional information of the BH spin parameter a further limitation can be done and is the next topic.

%%%%%%%%%%%%%%%%%%%%%%%%%%%%%%%%%%%%%%%%%%%%%%%
%  BH spin
%%%%%%%%%%%%%%%%%%%%%%%%%%%%%%%%%%%%%%%%%%%%%%%
%\subsection{Black hole spin}

As the BH is accreting mass from the donor, it is not only the BH's mass that is changing but also its angular momentum. We assume here that the material being accreted by the BH is carrying the specific angular momentum of the BH's innermost stable circular orbit (ISCO). For a BH with initial mass $M_{\mathrm{i}}$ and zero spin at the onset of RLO and total mass-energy $\mathcal{M}$ once some material has been accreted, where $1 \leq\tfrac{\mathcal{M}}{M_{\mathrm{i}}}\leq6^{\tfrac{1}{2}}$, the dimensionless Kerr spin parameter is 
\begin{equation}
\label{eq:BH_spin}
a_{\ast} = \left(\frac{2}{3}\right)^{\tfrac{1}{2}}\frac{M_i}{\mathcal{M}}\left[4-\sqrt{\frac{18M_{i}^2}{\mathcal{M}^2}-2}\right],
\end{equation}
for 0 $\leq a_{\mathrm{\ast}} \leq$ 1 \citep{thorne1974}.

The relation between the accreted rest mass $M_{\mathrm{0}} - M_{\mathrm{0i}}$ since RLO onset, the BH initial total mass-energy $M_{\mathrm{i}}$, and the total mass-energy $\mathcal{M}$ is
\begin{equation}
\label{eq:Maccret}
M_0 - M_{0i} = 3M_i\left[\sin^{-1}\left(\frac{\mathcal{M}}{3M_i}\right)-\sin^{-1}\left(\frac{1}{3}\right)\right].
\end{equation}
Here $M_{\mathrm{0i}} = 0$ assuming the BH has not accreted any mass prior to RLO onset and the total accreted rest mass is $M_{\mathrm{0}}$.
We will further assume that $a_{\mathrm{\ast}}$ = 0 at RLO onset, i.e. that the natal spin of the BH is negligible, which was shown to be a reasonable assumption for explaining the Galactic BH LMXB population \citep{2015ApJ...800...17F}. Under these assumptions, solving eq. \eqref{eq:Maccret} with respect to $\mathcal{M}$ and inserting it into eq. \eqref{eq:BH_spin} yields the BH spin parameter for an accreted amount of rest mass $M_{\mathrm{0}}$. Finally, for sake of completeness we also include the case $\tfrac{\mathcal{M}}{M_{\mathrm{i}}} \geq 6^{\tfrac{1}{2}}$ and $a_{\mathrm{\ast}} = 1$ where
\begin{equation}
\mathcal{M} = 3^{- \tfrac{1}{2}}M_0-3^{\tfrac{1}{2}}M_i \left[\sin^{-1}\left(\frac{2}{3}\right)^{\tfrac{1}{2}} - \sin^{-1}\left(\frac{1}{3}\right)\right] + 6^{\tfrac{1}{2}}M_i.
\end{equation}

\begin{figure}[!Htb]
\includegraphics[width=1.0\columnwidth]{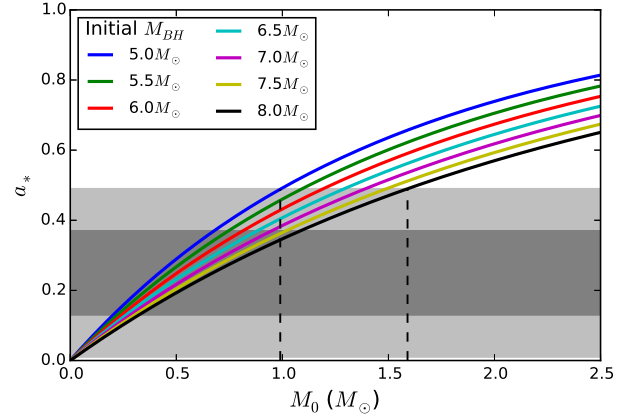}
\caption{The evolution of the BH spin parameter, given by eq. \eqref{eq:BH_spin}, for a set of BH initial masses $M_{\mathrm{i}}$, as a function of accreted mass $M_{\mathrm{0}}$ eq. \eqref{eq:Maccret}. The grey areas mark the 1$\sigma$ and 2$\sigma$ errors centered on the observed spin parameter for LMC X-3 (see Table \ref{tab:LMC_X-3}). The vertical dashed black lines indicate the maximum allowed accreted mass $M_{\mathrm{0}}$, at 2$\sigma$ limit, for an initial non-spinning BH mass of $5\,\rm \mathrm{M_{\odot}}$ (left) and $8\,\rm \mathrm{M_{\odot}}$ (right) respectively.}
\label{fig:BH_spin}
\end{figure}

The solution of eq. \eqref{eq:BH_spin} is shown in Fig. \ref{fig:BH_spin}, for different initial BH masses at RLO onset. Here the evolution of the BH spin parameter is tracked as a function of accreted mass. For reference, centered on the LMC X-3 spin parameter $a_{\mathrm{\ast}} = 0.25$, we have added the $1\sigma$ and $2\sigma$ error region as the dark and light gray areas respectively.
One notes how a small initial BH mass requires less accreted mass, in absolute terms, to build up its spin parameter compared to higher initial BH masses. 
% This makes sense since one would expect the inertia of an object to increase with its mass.
As an example we can follow a BH with initial mass $M_{\mathrm{i}} = 5.0 M_{\mathrm{\odot}}$ (blue line) as it accretes mass and compare it to a BH with initial mass $M_{\mathrm{i}} = 8.0 M_{\mathrm{\odot}}$ (black line). The blue line reaches the upper spin parameter having accreted $\sim1\mathrm{M_{\odot}}$ reaching a current mass of $6M_{\mathrm{\odot}}$. The black line reaches the upper spin parameter limit when it has accreted $\sim1.6M_{\mathrm{\odot}}$ and has a mass of $9.6 M_{\mathrm{\odot}}$.
In terms of relative mass accreted to initial BH mass both BHs accrete $\sim20\%$ of their initial BH mass before reaching the upper spin parameter limit of LMC X-3, so their initial BH mass was $\sim83.3\%$ of the current mass when accretion began. Hence, the maximum amount of accreted material onto the BH of LMC X-3 is $\sim16.6\%$ of its current mass. The LMC X-3 BH has a lower mass limit $M_{\mathrm{BH}} = 5.86M_{\mathrm{\odot}}$ within $2\sigma$. A lower limit of the BH mass at RLO onset therefore becomes $(1-0.166)5.86\mathrm{M_{\odot}}\sim4.9M_{\mathrm{\odot}}$.

%In the above example, we considered the spin parameter change through accretion together with the lower $M_{\mathrm{BH}}$ mass constraint which gave a lower limit of the BH mass at RLO onset. 
If we instead consider the observed mass of the LMC X-3 BH and accept a $2\sigma$ limit, its current upper mass limit is $8.1M_{\mathrm{\odot}}$. Thus, a BH of initial mass $8.0 M_{\mathrm{\odot}}$ cannot have accreted more than $0.1M_{\mathrm{\odot}}$ suggesting such a BH in the case of LMC X-3 should have a spin parameter $a_{\mathrm{\ast}}\sim0.04$.

We stress here, we are using the observed BH spin of LMC X-3 only as a condition to constrain the \textit{upper limit} of the amount of mass that the BH can possibly have accreted since the onset of the RLO. In reality, the BH in LMC X-3 may have been formed with a non-zero natal spin, in which case the limits on the maximum accreted mass discussed above would be upper bounds.

%Setting an upper limit for the BH mass at RLO onset is more tricky because we can only guess what is the origin of the BH spin, either it is natal or from accretion. If we assume some part of or all the spin is natal the upper limit of the BH mass at RLO onset $M_{BH,RLO} \rightarrow 8.1M_{\mathrm{\odot}}$ as $a_{\ast,RLO} \rightarrow 0.49$. In the extreme case with $M_{BH,RLO} = 8.1M_{\mathrm{\odot}}$ and $a_{\ast,RLO} = 0.49$ is the upper limit to the BH mass at RLO onset. Also in the extreme case the mass loss fraction in the vicinity of the BH $\beta = 1$ as the donor star cannot transfer any mass towards the BH without the mass being lost and it would give a new solution instead of eq. \eqref{eq:orbit_evo} for the orbital evolution.

%Possible solutions of eq. \eqref{eq:orbit_evo} are presented as white areas in figure \ref{fig:MTseq_no_Edd} and is discussed later in this section. The analytical point-mass MT model is an effective method to reduce the potential solution space for the considered parameters. With the additional information of the BH spin parameter a further limitation can be done and is the next topic.

In summary, we note that the analytic point-mass MT model together with the BH spin parameter is a powerful yet computational inexpensive tool to establish rough but meaningful bounds on our 5-dimensional parameter space in determining whether a set of initial conditions at RLO onset lead to a MT sequence that satisfies the observational constraints on $P_{\mathrm{obs}}$, $M_{\mathrm{BH}}$, and $M_{\mathrm{2}}$. To account for the remaining observational constraints, X-ray luminosity, $T_{\mathrm{eff}}$, $R_{\mathrm{2}}$, $\log g_{\mathrm{2}}$, and donor luminosity $L_{\mathrm{2}}$, we need to model in detail the stellar evolution of the companion star.

%%%%%%%%%%%%%%%%%%%%%%%%%%%%%%%%%%%%%%%%%%%%%%%
%  MESA detailed simulation
%%%%%%%%%%%%%%%%%%%%%%%%%%%%%%%%%%%%%%%%%%%%%%%
\subsection{Detailed MESA calculation}\label{sec:setup:mesa}

Our detailed simulations of the MT phase between the donor star and the BH are done with MESA, a state-of-the-art, publicly available, 1-D stellar structure and binary evolution code. We adopted a metallicity for the donor star appropriate for LMC stars \citep[Z=0.006; which is in line with derived metallicities of stellar populations in LMC][]{Antoniou:2016iw} and used the implicit MT scheme incorporated in MESA. All our simulations were done with revision 7184 of the code\footnote{The detailed MESA settings used in out simulations can be found at http://mesastar.org/results}. We explored the initial parameter space at the onset of RLO by constructing a grid of MT sequences for $M_{\mathrm{2, RLO}}$ in the range 2-15 $M_{\mathrm{\odot}}$ in steps of 0.2 $M_{\mathrm{\odot}}$, orbital periods $P_{\mathrm{RLO}}$ from 0.6-4.2 days in steps 0.1 day, and $M_{\mathrm{BH, RLO}}$ from 5.0-8.0 $M_{\mathrm{\odot}}$ in steps of 0.5 $M_{\mathrm{\odot}}$. Furthermore, we vary the accretion efficiency parameter $\beta$, considering values of $\beta$=0.0,\,0.25,\, 0.5,\, 0.75,\, 0.9, and 1.0. In all our calculations we set the accretion efficiency parameter $\alpha$ = 0.0, as a non-zero value would be appropriate only for massive donor stars that lose a significant amount of mass in the form of fast stellar winds \citep{1994inbi.conf..263V}. To speed up the computation of individual models we have invoked a stopping criterion within MESA which terminates the computation once the orbital period is above 6 days. Out of all possible combinations of $M_{\mathrm{2, RLO}}$, $M_{\mathrm{BH, RLO}}$, $P_{\mathrm{RLO}}$ and $\beta$, we only run detailed MESA calculations for those sets of initial conditions that ``passed'' our analytic MT model test, described in Sect. \ref{sec:analytic_model}.

% As matter is accreted onto a compact object potential energy is radiated away at the moment of accretion. The Eddington accretion limit is the pressure of the in-falling material being equaled by the pressure of the outgoing radiation created from the release of potential energy as matter is being accreted onto the BH. Assuming the interaction between radiation and in-falling matter is due to proton electron scattering and the matter consist of hydrogen the Thompson scattering cross-section applies and the Eddington accretion limit $\dot{M}_{Edd}$ is \citep{frank2002}
% \begin{equation}
% \dot{M}_{Edd} = \frac{L_{Edd}}{\eta c^2}
% \label{eq:dMedd}
% \end{equation}
% where L$_{Edd}$ is the standard Eddington luminosity, $\eta$ = 0.1 is the rest mass to radiation efficiency, and c is the speed of light in vacuum.
% In order to examine the effect of the Eddington accretion limit eq. \eqref{eq:dMedd} on the evolution of the system, we ran two sets of MT sequences. One applying the Eddington accretion limit and one without this limit.

%%%%%%%%%%%%%%%%%%%%%%%%%%%%%%%%%%%%%%%%%%%%%%%
%  Search each simulation
%%%%%%%%%%%%%%%%%%%%%%%%%%%%%%%%%%%%%%%%%%%%%%%
To identify those MT sequences that are PPS of LMC X-3 at the moment of RLO onset we perform the following checks:

\begin{enumerate}
%%%%%%%%%%%% 
\item The ZAMS radius of the donor star is smaller than its Roche lobe radius at the onset of RLO.

%%%%%%%%%%%% 
\item When RLO begins the donor star's age is less than the age of the Universe (<13.7\,Gyr). Furthermore, the MT calculations are terminated when the age of the donor reaches 13.7\,Gyr, so the system today cannot be older than the age of the Universe either. In practice, neither of these constraints limit our calculations here, due to the range of donor star masses that we are exploring.

%%%%%%%%%%%% 
\item During the MT phase, as the orbital period is evolving, it should cross the observed period $P_{\mathrm{obs}}$. 
% such that during a set of consecutive and sufficiently small time steps $\Delta t$ the calculated orbital period P fulfils the inequality
% \begin{equation}
% P(t) < P_{obs} < P(t + \Delta t)
% \end{equation}
% and we assume that either P(t) or P(t+$\Delta$t) is sufficiently close to the observed period that they are identical.

%%%%%%%%%%%% 
\item A MT sequence model $\mathcal{I}$ has $j = [1,5]$ predicted parameters which all must be within 2$\sigma$ of the observed quantities $\mu_{\mathrm{j}}$ given their associated errors $\sigma_{\mathrm{j}}$ both listed in Table \ref{tab:LMC_X-3}
\begin{equation}
\label{eq:accept_criteria_4}
\frac{|\mathcal{I}_{i,j} - \mu_j)|}{\sigma_j} \leq 2
\end{equation}
where $\mathcal{I}_{\mathrm{i,j}}$ are the predictions of the $i^{\mathrm{th}}$ MT sequence for each of these quantities, at the time when the orbital period of the MT sequence is equal to the observed one.
%%%%%%%%%%%% 
\item Infer whether the predicted accretion rate onto the BH by the specific MT system would cause it to appear as a transient or a persistent X-ray source.

\end{enumerate}

%Check 1 is necessary since a collapsing cloud or proto-star between the pre-stellar phase and ZAMS cannot fill its Roche lobe without it affecting the end product of that star formation. We perform check 2 for consistency. Chechk 3 is necessary to avoid simulating a system that matches the 6 observationally inferred parameters of Table \ref{tab:LMC_X-3} but do not match the observed orbital period. With check 3 we also avoid accounting for any interpolation issues from the simulated MT sequences to the observed orbital period. In check 4 we use a 2$\sigma$ error margin to get a conservative estimate of allowed ranges. 
The latter check separates the MT sequences into two regimes of MT mechanisms. If the accretion rate onto the BH is $\dot{M}_{\mathrm{BH}}$ < $\dot{M}_{\mathrm{BH,crit}}$, where
\begin{equation}
\label{eq:accept_criteria_5}
\dot{M}_{\mathrm{BH,crit}} \approx 10^{-5} \left( \frac{M_{\mathrm{BH}}}{\mathrm{M_{\odot}}} \right)^{0.5} \left( \frac{M_{\mathrm{2}}}{\mathrm{M_{\odot}}} \right)^{-0.2} \left( \frac{P}{1\mathrm{yr}} \right)^{1.4} \frac{\mathrm{M_{\odot}}}{\mathrm{yr}},
\end{equation} 
the system is in a transient state, otherwise the system is in a persistent state \citep{dubus1999}. A persistent source has a continuous flow of material from the donor star to the BH, which allows for a relatively high luminosity. If the source is transient, it goes through periods of outbursts of high X-ray flux but spends most of its time with little to no X-ray flux, i.e. in a quiescent phase.

Under the assumption of constant MT efficiency, i.e. constant $\alpha$ and $\beta$, MT sequences passing checks 1 through 4 are PPS that predict a system with the characteristics similar to currently observed properties of LMC X-3. In principle $\alpha$ and $\beta$ are average values for the efficiency of the MT and therefore merely serve as an indicator of whether a semi-conservative or non-conservative transfer of mass is needed.

Assuming that the reported errors for all observed quantities in Table \ref{tab:LMC_X-3} are Gaussian, then the likelihood of a model MT sequence, $\mathcal{I}_{\mathrm{i}}$, given the currently observed data, $D$, of LMC X-3 can be written as:
\begin{equation}
\label{eq:likelihood}
\mathcal{L}_i(\mathcal{I}_i|D) = \prod\limits_{j = 1,5} {\mathcal{G}({\mathcal{I}_{i,j}};\mu_j,\sigma _j)},
\end{equation}
where $\mathcal{G}$ is a normalised Gaussian function, and $\mu_{\mathrm{j}}$ and $\sigma_{\mathrm{j}}$ for $j = [1,5]$ are the observed values and the associated errors listed in Table \ref{tab:LMC_X-3}. In principle, the likelihood of a MT sequence, when the sequence is crossing the observed period, is an estimate of how close the specific model MT sequence comes to the observed properties of LMC X-3. We intend to use eq. \eqref{eq:likelihood} as a weight to produce a series of probability density functions (PDF) in Sect. \ref{sec:BSE}.
In this case, we also multiply the likelihood value calculated by eq. \eqref{eq:likelihood} by the time period, $\Delta T$, that a model MT sequence spent close to the observed orbital period. This time period is defined as the total time during which all parameters $j$ are within 2$\sigma$ of the observed properties.

% Individual runs %%%%%%%%%%%%%%%%%%%%%%%%%%%%%%%%%%%%%%%%%%%%
%1) Describe individual runs to illustrate how the locate and search function works for a winner model works.
\subsection{Constraining the Properties of LMC X-3 progenitor at the Onset of Roche Lobe Overflow}
With the checks defined in the previous section we now present three examples of individual MT sequences and compare these to the observational parameters in Table \ref{tab:LMC_X-3}. The examples are chosen to demonstrate how the constraints select some MT sequences as PPS and discard others.

\begin{figure*}[!Htb]
\raggedleft
\includegraphics[width=1.0\linewidth]{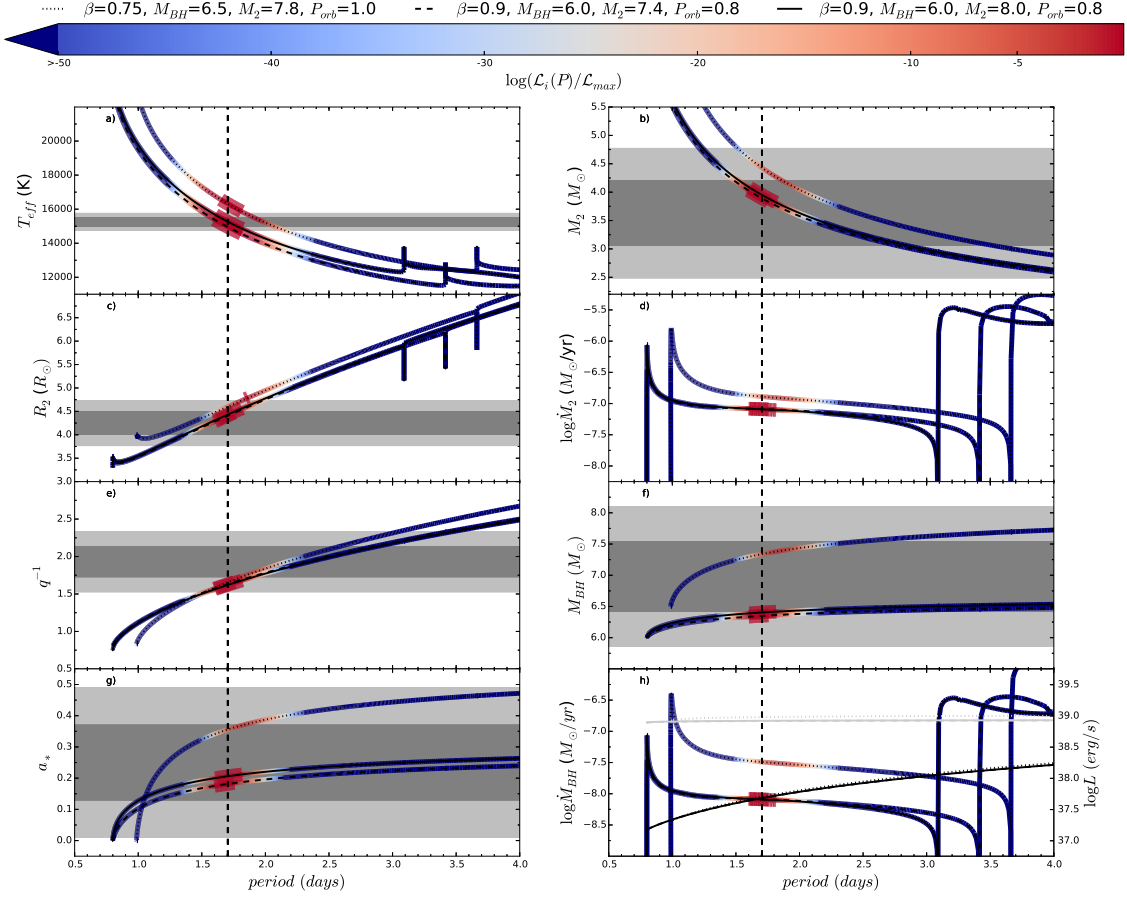}
\caption{The solid, dashed and dotted lines represent examples of a binary system undergoing RLO and MT for different sets of initial values at the onset of RLO. The vertical dashed line is the observed orbital period given in Table \ref{tab:LMC_X-3}. The color map indicates the relative likelihood $\log(\mathcal{L}_{\mathrm{i}}/\mathcal{L}_{\mathrm{max}})$ from eq. \eqref{eq:likelihood}. The thickening of the color bands around the observed orbital period, denote that in this region all model parameters $\mathcal{I}_{i,j}$, different from the one being plotted, are within 2$\sigma$ of the observed value from Table \ref{tab:LMC_X-3}. In all panels but (f) and (h), the dark and light grey regions are centered on the observed value and each grey contrast extends to an area of $\pm 1\sigma$ and $\pm 2\sigma$ respectively. The black lines in panel h) are the critical accretion rate eq. \eqref{eq:accept_criteria_5} and the light-grey lines are the Eddington accretion rate \citep{frank2002}. We note that all three examples displays first a case A MT where core hydrogen takes place. The sudden changes seen at periods equal to 3.1, 3.4, and 3.7 days, respectively, is the transition from case A MT, to case B MT, where the donor star has formed a helium core and hydrogen-shell burning has commenced. Computation of the three examples shown are stopped before case B MT is completed, when the orbital period reaches 6 days.}
\label{fig:MTseq}
\end{figure*}

All three examples are plotted together in Fig. \ref{fig:MTseq} and have initial parameters at the onset of RLO: $\beta$=0.75, $M_{\mathrm{BH, RLO}}=6.5M_{\mathrm{\odot}}$, $M_{\mathrm{2, RLO}}=7.8M_{\mathrm{\odot}}$, and $P_{\mathrm{RLO}}=1.0$days  (dotted lines); $\beta$=0.90, $M_{\mathrm{BH, RLO}}=6.0M_{\mathrm{\odot}}$, $M_{\mathrm{2, RLO}}=7.4M_{\mathrm{\odot}}$, and $P_{\mathrm{RLO}}=0.8$days  (dashed lines); $\beta$=0.90, $M_{\mathrm{BH, RLO}}=6.0M_{\mathrm{\odot}}$, $M_{\mathrm{2, RLO}}=8.0M_{\mathrm{\odot}}$, and $P_{\mathrm{RLO}}=0.8$days  (solid lines).
Along the x-axis of all panels is the orbital period $P$ in days and the vertical dashed line marks the currently observed orbital period $P_{\mathrm{orb}}$ (see Table \ref{tab:LMC_X-3}).
The color bar shows the relative likelihood $\mathcal{L}_{\mathrm{i}}/\mathcal{L}_{\mathrm{max}}$, as determined from eq. \eqref{eq:likelihood} for all parameters $j$. For all three examples, a maximum likelihood is reached as the MT sequence model crosses the observed orbital period.
In each panel, (a) through (h), a thickening of the color indicates that all parameters $\mathcal{I}_{\mathrm{i,j}}$ apart from the one being plotted, as indicated on each y-axis, are within 2$\sigma$ of the observed value. The grey areas shown in some of the panels indicate the 1$\sigma$ and 2$\sigma$ (dark and light grey, respectively) error regions of those parameters centered on the observed values, similar to Fig. \ref{fig:BH_spin}.

%%%%% T_eff with Period %%%%%
Panel \textbf{(a)} of Figure \ref{fig:MTseq} shows the evolution of the donor star's effective temperature with orbital period. The solid and dashed lines are within 2$\sigma$ of $T_{\mathrm{eff,obs}}$ as they cross the observed orbital period and are thickened, indicating that all other model parameters $j$ are also within the 2$\sigma$ region of their observed values. The solid and dashed lines are likewise thickened on all panels in Fig. \ref{fig:MTseq}, close to the observed orbital period. The dotted line is thickened in panel \textbf{(a)} and \textbf{(b)}, but falls outside the effective temperature 2$\sigma$ region and therefore is not thickened on the other panels of Fig. \ref{fig:MTseq}, indicating that the specific MT sequence does not represent a PPS. Qualitatively, the effective temperature decreases until the orbital period reached 3.1, 3.4, and 3.7 days, respectively, where it briefly spikes, signaling the end of the donor's main sequence. Following the spike, the effective temperature continues its decrease, though it has experienced a slight increase relative to that before the spike.

%%%%% M2 with Period %%%%%
Panel \textbf{(b)} of Figure \ref{fig:MTseq} shows the donor's mass, $M_{\mathrm{2}}$, as a function of the orbital period. The donor mass $M_{\mathrm{2}}$ is not experiencing any changes around orbital periods of 3.1, 3.4, and 3.7 days respectively, indicating that the changes observed in temperature is due to the donor star's internal structure and not the result of sudden changes in the secular evolution of the binary system. We remind that the donor mass is not used in deriving the relative likelihoods nor is it included as constraining parameter, instead the mass ratio $q$ is.

%%%%% R2 with Period %%%%%
The donor's radius $R_{\mathrm{2}}$, shown in panel \textbf{(c)}, increases as the orbital period increases. As the star evolves away from the MS, the donor star radius increases to the Roche lobe limit, causing a mass loss. With this mass loss, the orbit expands, the Roche lobe radius increases, and the donor star radius can further increase. This process continues until the donor star reaches the end of H burning where the donor star radius shows a contraction at orbital periods of 3.1, 3.4, and 3.7 days. Following the contraction, the donor star expands to fill its Roche lobe again. During the MS the donor radius increases due to its core molecular weight increasing and results in a more dense and luminous core, causing the above layers to expand \citep{Maeder2008}. With the decrease in mass and increase in radius of the donor star, its surface gravity, not shown, also decreases, similar to the effective temperature.

%%%%% Mdot_{\mathrm{2}} with Period %%%%%
In contrast to the mass of the donor star monotonically decreasing, the evolution of mass loss rate $\dot{M_{\mathrm{2}}}$, shown in panel \textbf{(d)} of Fig. \ref{fig:MTseq} varies widely. It begins with a rapid turn-on phase, when the donor star fills its Roche lobe, and then gradually declines as the mass-ratio of the binary decreases.
%It is seen that the first rapid increases are just at the initiation of the simulations which indicates the turn on of the RLO. 
The initial turn-on phase rapidly reaches a peak MT rate near $\mathrm{10^{-6}\,M_{\odot}\,yr^{-1}}$ and then drops by nearly an order of magnitude, where it stabilises to a steady slow decrease until an orbital period of 3.1 days for the solid line, 3.4 days for the dashed, and 3.7 days for the dotted lines. At this point, the rate drops below $\mathrm{10^{-10}\,M_{\odot}\,yr^{-1}}$ and the binary temporarily detaches, but MT commences again with a rate of $\sim \mathrm{10^{-5.5}\,M_{\odot}\,yr^{-1}}$, which is roughly sustained until the whole envelope of the donor star is removed (see also Sect. \ref{sec:1915} where the evolution of LMC X-3 after its currently observed state is described).
The mass loss rate of the donor star $M_{\mathrm{2}}$ qualitatively dictates the accretion rate onto the BH shown in panel \textbf{(h)} of Fig. \ref{fig:MTseq} but is larger by a factor $(1-\beta)^{\mathrm{-1}}$·
%Again, the changes in the mass loss rate happens at the same time as changes in the effective temperature and radius. Comparing the two turn-on episodes, one sees, that the force driving the donor star envelope outwards is larger than the force that initiated the RLO to begin with.

%%%%% log g with Period %%%%%
%With the decrease in mass of the donor star, its surface gravity decreases and like the effective temperature the surface gravity also experiences a spike at orbital periods 3.1 (dashed) and 3.7 days (dash dotted and solid) respectively. All three lines have nearly similar values for the surface gravity at the crossing period and are within the error region.
Panel \textbf{(e)} of Fig. \ref{fig:MTseq} shows the change in the systems mass ratio which increases as material is transferred to the BH or is lost from the system. For the three examples shown, both their mass ratio and donor mass at the orbital period is within 2$\sigma$. However, there are combinations of donor mass and BH mass for which the mass ratio is outside the 2$\sigma$ region. 

%%%%% Mbh and BH spin with Period %%%%%
In panel \textbf{(f)} of Fig. \ref{fig:MTseq}, we show the evolution of the BH's mass as a function of orbital period. The accretion onto the BH is initially mildly super-Eddington, but declines rapidly. Especially for the dashed and solid lines with $\beta$=0.9, it is seen that the BH mass-accretion rate reaches a plateau with only a small gain in mass as the orbital period lengthens. For the solid line with $\beta$=0.75 the BH mass increase is larger, while at the same time the orbit expands faster. All three lines satisfy the observed BH mass constraint at the crossing period. The BH spin $a_{\ast}$, shown in panel \textbf{(g)}, is a measure of the accreted angular momentum by the BH, as we assumed a zero natal BH spin. We find that the evolution of the BH spin follows the characteristics of the BH mass, i.e. the growth in spin is steep initially and flattens as the orbital period increases.

%%%%% Mdot_BH with Period %%%%%
Finally, panel \textbf{(h)} shows the BH mass accretion rate as a function of orbital period. The light grey lines are the Eddington accretion rates \citep{frank2002} and in black are the critical accretion rates (eq. \eqref{eq:Maccret}), that separate transient from persistent behavior based on the thermal disk instability model, for each MT sequence. The changes in the BH mass accretion rate with orbital period are similar to the mass loss rate of the companion star, however, scaled with the value close to $\beta$ which is the fraction of transferred mass that is lost from the system in the vicinity of the BH. The solid line predicts a system which is in the transient state, though not until an orbital period above $\sim$1.65 days. Both the dashed and dotted lines are predicted to be persistent systems when they cross the observed orbital period. The BH accretion rate is also affected by the companion star's sudden change in radius and effective temperature, at orbital periods of 3.1, 3.4, and 3.7 days, respectively. Following the second MT turn-on event, the BH in the three example sequences accrete close to the Eddington accretion rate and are potentially mildly super-Eddington.

The three examples shown in Fig. \ref{fig:MTseq} all show the same qualitative evolution, but due to the differences in their initial conditions display the changes at different orbital periods.
The radial contraction seen in symphony with the increase in effective temperature and surface acceleration, at 3.1, 3.4, and 3.7 days, indicates that the donor star is restructuring its interior. Its core has depleted its hydrogen, stopped nuclear burning, and left behind a helium core. As a result, the core luminosity briefly drops, whereby the envelope contracts, i.e. the donor radius decreases, and the effective temperature and surface acceleration increase. As a consequence of the contraction, a shell of hydrogen at the bottom of the envelope ignites, pushing the envelope beyond its pre-contraction radius. Hence the three examples in Fig. \ref{fig:MTseq} are case A RLO MT until an orbital period of 3.1, 3.4, and 3.7 days, where the hydrogen shell burning forces the envelope outwards and the three example enter a RLO MT of case B \citep{kippenhahn1967}.
In single star evolution, a star entering the hydrogen shell burning increases its radius dramatically, by approximately two orders of magnitude \citep{tauris2006}, and the star enters the Red Giant Branch (RGB). But due to the presence of the BH, the RLO prevents the donor star to evolve into the RGB phase. The donor star on one hand loses mass, increasing its Roche limit, and on the other hand inflates due to its secular evolution which will shrink its Roche limit. Near shell burning ignition the donor star has lost $\sim$60$\%$ of its ZAMS mass. Due to our stop criteria at an orbital period of 6 days, the three examples shown in Fig. \ref{fig:MTseq} are not entering the helium burning phase. Though not shown, the rate of change in orbital period during case A MT is slow and during the case B MT phase increases dramatically, such that the entire evolution shown during case B MT in Fig. \ref{fig:MTseq} only takes a few $\mathrm{10^5}$ yr. The donor star, in our three examples, has lost as much as $\sim$70$\%$ of its ZAMS mass by the time our simulations stop. For our MT sequences we expect the donor star's stellar evolution to end with the formation of a He White Dwarf (WD) or in the cases of the most massive donors a Carbon Oxygen WD.

%%%%% Rounding off MT sequence examples from MESA %%%%%%
By reviewing the evolution of three selected MT sequence models, one that passes items 1-3 (dotted line), defined in Sect. \ref{sec:setup:mesa}, two that pass items 1-4, we find confidence in our checks as a filter to locate PPS of LMC X-3 at the moment of RLO onset. As we also show, we can discriminate between transient and persistent systems. The calculated relative likelihood shows a maximum for all three examples close to the orbital period, a validation of our approach. We next look at the overall result of our detailed MT sequences.

%%%%% All MESA MTseq %%%%%
\subsection{Results of modelling the X-ray binary phase}\label{sec:MTseq_results}

\begin{sidewaysfigure*}%[htb]
%\sidecaption
\centering
\includegraphics[width=1.0\hsize]{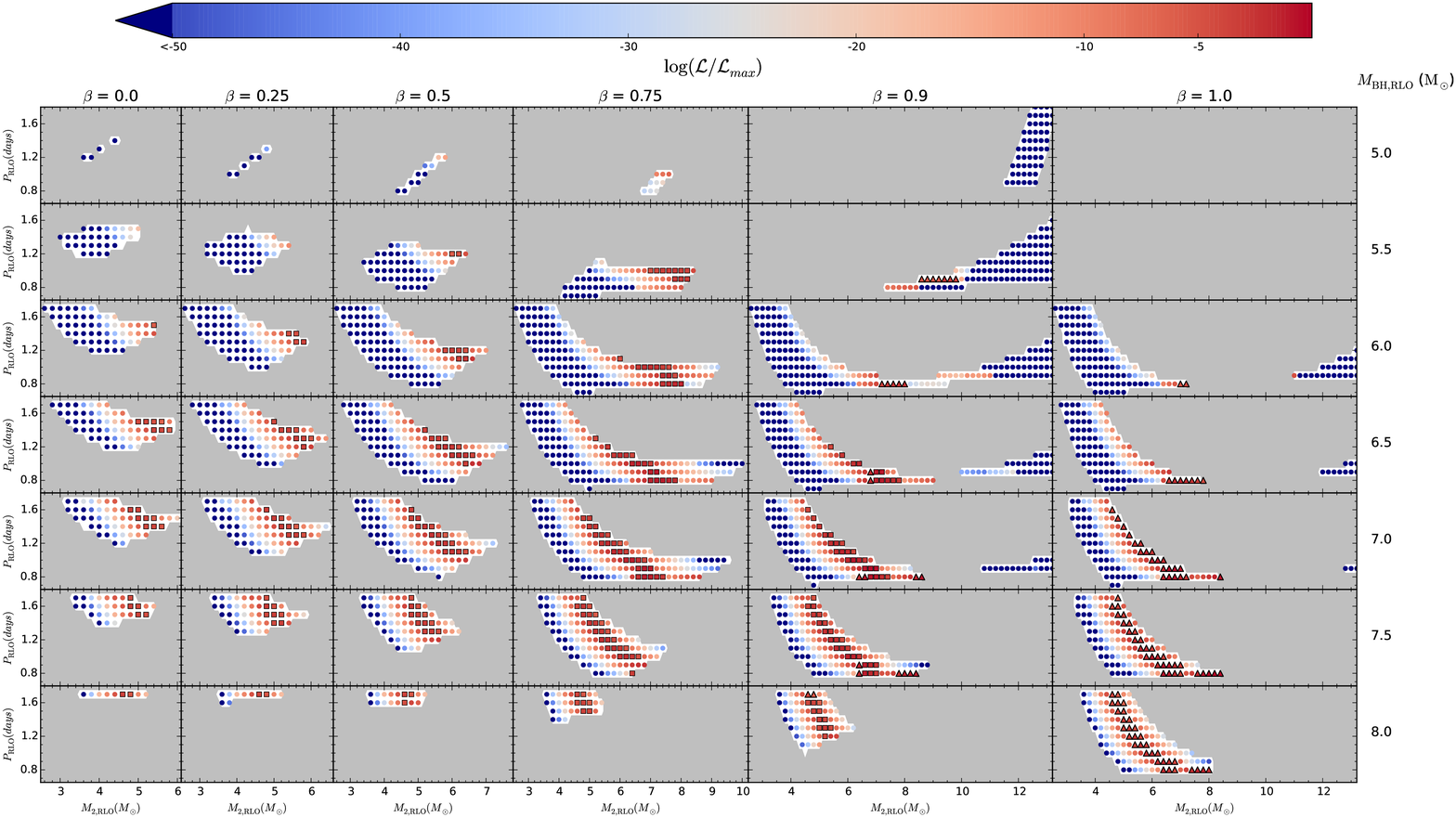}
\caption{Grid of MT sequences for different MT efficiencies of $\alpha$=0.0 and $\beta$=[0.0, 0.25, 0.5, 0.75, 0.90, 1.0], going from the left to the right columns of panels. Each row, as labeled on the right of the figure, corresponds to an initial value of $M_{\mathrm{BH, RLO}}$=[5.0, 5.5, 6.0, 6.5, 7.0, 7.5, 8.0]$\,M_{\mathrm{\odot}}$. On the x-axis of each panel are shown the initial mass of the donor star $M_{\mathrm{2, RLO}}$, the range of which is allowed to vary with $\beta$. The y-axis shows the initial orbital period $P_{\mathrm{RLO}}$ in days. Grey areas corresponds to regions with initial $M_{\mathrm{BH, RLO}}$, $M_{\mathrm{2, RLO}}$, $P_{\mathrm{RLO}}$ values which do not pass the analytic MT test or the initial ZAMS radius is greater than the Roche lobe. The white areas represent initial combinations of $M_{\mathrm{BH, RLO}}$, $M_{\mathrm{2, RLO}}$, $P_{\mathrm{RLO}}$ that pass the analytic MT model, and for which we have made a detailed simulation with MESA. Circles indicates MT sequences which fail check 4, squares pass all checks and are persistent solutions according to eq. \eqref{eq:accept_criteria_5}, and triangles are transient solutions and passes all checks. The colors indicate the relative likelihood of each detailed run.}
\label{fig:MTseq_no_Edd}
\end{sidewaysfigure*}

In Fig. \ref{fig:MTseq_no_Edd}, initial conditions of our grid of 3319 MT sequences are shown. Each column of panels represents MT sequences for different values of $\beta$ (top axis). Each row varies the initial value of $M_{\mathrm{BH, RLO}}$ in units of $M_{\mathrm{\odot}}$ (right axis). On the x-axis within each panel are plotted the companion mass $M_{\mathrm{2, RLO}}$ in units of $M_{\mathrm{\odot}}$ and on the y-axis the orbital period $P_{\mathrm{RLO}}$ in days. The grey areas indicate regions where any combination of initial values fails the analytic MT model. The white areas denote combinations of initial values which pass the analytic MT model. As one increases $\beta$, the relative size of the area passing the analytic model increases. As a result we allow the horizontal size of each panel and x-axis range to change with $\beta$ in order to better display the content of each figure.

Each circle, square, and triangle represents a set of initial conditions of a binary system right before initiation of RLO that was simulated with MESA. Circles are those MT sequences which fail check 4. Those sequences that pass all checks and are persistent solutions are shown as squares, while triangles denote those sequences that pass all checks and are transient solutions. It is clearly shown in the figure that the number of PPS increases with $\beta$. Whereas persistent solutions (squares) can be found for all values of $\beta$ except $\beta$=1.0, transient solutions are only found for $\beta$=0.9 and $\beta$=1.0, indicating that if LMC X-3 is a transient source the ongoing MT is highly non-conservative.

The color represents the relative likelihood of a particular MT sequence model to the model with maximum weight, similar to that used for Fig. \ref{fig:MTseq}, displaying individual MT sequences. Note that while in Fig. \ref{fig:MTseq} we show the relative likelihood as a function of orbital period, in Fig. \ref{fig:MTseq_no_Edd} we display the value of the relative likelihood of each MT sequence at the observed orbital period.
The set of initial conditions with the highest likelihood is $\beta$=1.0, $M_{\mathrm{BH,RLO}}$ = 7.5$\,\mathrm{M_{\odot}}$, $M_{\mathrm{2,RLO}}$ = 8.2$\,\mathrm{M_{\odot}}$, and $P_{\mathrm{RLO}}$ = 0.8 days which is a persistent solution. Of the 3319 MT sequences simulated with MESA, 395 are PPS of LMC X-3 at the moment of RLO onset. Of these, 274 are persistent solutions and 121 are transient solutions.

%\newgeometry{left=1cm,right=1cm,top=2.0cm,bottom=1.0cm}
%\begin{landscape}
%\end{landscape}
%\restoregeometry

%\newgeometry{left=1cm,right=1cm,top=2.0cm,bottom=1.0cm}
%\begin{landscape}
\begin{sidewaystable*}
\centering
\caption{Selected properties of some MT sequences that pass all criteria at the observed orbital period. The 10 MT sequences numbered between 175 to 206 are persistent solutions and the 10 MT sequences numbered between 1970 to 2224 are transient solutions. For all MT sequence models $\alpha = 0$. A full list of all MT sequences found is available in the on line version.}
\label{tab:set_winners}
\begin{tabular}{cccccccccccccccc}
\hline
\hline
\rule{0pt}{2.5ex} & \multicolumn{7}{c}{Parameters at RLO onset} & &  \multicolumn{6}{c}{Parameters at observed orbital period $P_{\mathrm{orb}}$}\\ %[0.5ex]
\cline{2-7} \cline{9-15}
%\rule{0pt}{2.5ex} 1 & 2 & 3 & 4 & 5 & 6 & 7 & 8 & 9 & 10 & 11 & 12 & 13 & 14 & 15 \\ 
\makecell{MT \\ Sequence} & ${\beta}$\tablefootmark{a} & $M_{\mathrm{BH, RLO}}$\tablefootmark{b} & $M_{\mathrm{2, RLO}}$\tablefootmark{c} & $P_{\mathrm{RLO}}$\tablefootmark{d}  & $X_{\mathrm{c, RLO}}$\tablefootmark{e} & $\tau_{\mathrm{RLO}}$\tablefootmark{f} & & $M_{\mathrm{BH}}$\tablefootmark{b} & $a_{\ast}$\tablefootmark{g} & $M_{\mathrm{2}}$\tablefootmark{c} & $\log L_{\mathrm{2}}$\tablefootmark{h} & $X_{\mathrm{c}}$\tablefootmark{i} & $\tau$ & $\log(\mathcal{L}/\mathcal{L}_{\mathrm{max}})$\tablefootmark{j} & $\Delta T$\tablefootmark{k} \\

 & & ($M_{\mathrm{\odot}}$) & ($M_{\mathrm{\odot}}$) & (days) & & (Myr) &  & ($M_{\mathrm{\odot}}$) & & ($M_{\mathrm{\odot}}$) & ($L_{\mathrm{\odot}}$) & & (Myr)  & & (Myr) \\
\hline

175 & 0.0 & 8.0 & 4.6 & 1.7 & 0.12 & 86.26 &   & 8.02 & 0.01 & 4.57 & 3.00 & 0.09 & 87.91 & -6.98 & 0.52\\
188 & 0.0 & 7.0 & 4.8 & 1.6 & 0.14 & 77.56 &   & 7.29 & 0.13 & 4.50 & 2.99 & 0.12 & 78.62 & -6.75 & 0.13\\
190 & 0.0 & 7.5 & 4.8 & 1.6 & 0.18 & 74.75 &   & 7.80 & 0.13 & 4.49 & 2.99 & 0.12 & 78.52 & -6.71 & 0.38\\
191 & 0.0 & 7.5 & 4.8 & 1.7 & 0.14 & 77.56 &   & 7.55 & 0.02 & 4.74 & 3.06 & 0.10 & 79.77 & -7.72 & 0.34\\
192 & 0.0 & 8.0 & 4.8 & 1.7 & 0.14 & 77.56 &   & 8.04 & 0.02 & 4.75 & 3.06 & 0.10 & 79.65 & -8.55 & 0.35\\
200 & 0.0 & 6.5 & 5.0 & 1.5 & 0.18 & 68.30 &   & 7.14 & 0.28 & 4.35 & 2.96 & 0.15 & 70.16 & -7.62 & 0.10\\
203 & 0.0 & 7.0 & 5.0 & 1.5 & 0.18 & 68.30 &   & 7.60 & 0.25 & 4.39 & 2.97 & 0.15 & 70.38 & -6.80 & 0.40\\
204 & 0.0 & 7.0 & 5.0 & 1.6 & 0.18 & 68.30 &   & 7.36 & 0.16 & 4.63 & 3.03 & 0.13 & 71.47 & -7.24 & 0.20\\
205 & 0.0 & 7.5 & 5.0 & 1.5 & 0.18 & 68.30 &   & 8.04 & 0.22 & 4.45 & 2.99 & 0.15 & 70.52 & -7.04 & 0.41\\
206 & 0.0 & 7.5 & 5.0 & 1.6 & 0.18 & 68.30 &   & 7.82 & 0.14 & 4.67 & 3.05 & 0.13 & 71.47 & -7.64 & 0.27\\
&&&&&&&&&&&&&&& \\ 
1970 & 0.9 & 8.0 & 4.6 & 1.7 & 0.12 & 86.26 &   & 8.00 & 0.00 & 4.58 & 3.01 & 0.09 & 87.91 & -6.94 & 6.01\\
2010 & 0.9 & 8.0 & 4.8 & 1.7 & 0.14 & 77.56 &   & 8.00 & 0.00 & 4.77 & 3.07 & 0.10 & 79.66 & -8.82 & 0.36\\
2178 & 0.9 & 7.0 & 6.4 & 0.8 & 0.58 & 15.60 &   & 7.24 & 0.11 & 3.98 & 2.92 & 0.22 & 48.42 & -5.33 & 1.31\\
2181 & 0.9 & 7.5 & 6.4 & 0.8 & 0.58 & 15.60 &   & 7.73 & 0.10 & 4.08 & 2.95 & 0.22 & 47.86 & -4.79 & 1.75\\
2182 & 0.9 & 7.5 & 6.4 & 0.9 & 0.50 & 22.50 &   & 7.73 & 0.10 & 4.05 & 2.95 & 0.22 & 45.98 & -4.69 & 1.79\\
2189 & 0.9 & 7.0 & 6.6 & 0.8 & 0.58 & 14.46 &   & 7.25 & 0.12 & 4.05 & 2.95 & 0.22 & 45.89 & -4.27 & 1.91\\
2198 & 0.9 & 6.5 & 6.8 & 0.8 & 0.58 & 13.47 &   & 6.78 & 0.14 & 3.97 & 2.94 & 0.23 & 44.59 & -5.05 & 1.87\\
2199 & 0.9 & 6.5 & 6.8 & 0.9 & 0.51 & 19.44 &   & 6.79 & 0.14 & 3.91 & 2.93 & 0.24 & 42.64 & -5.01 & 1.86\\
2215 & 0.9 & 6.0 & 7.2 & 0.8 & 0.58 & 11.83 &   & 6.33 & 0.17 & 3.86 & 2.92 & 0.23 & 42.83 & -6.66 & 2.00\\
2224 & 0.9 & 6.0 & 7.4 & 0.8 & 0.60 & 9.56 &   & 6.35 & 0.18 & 3.89 & 2.94 & 0.23 & 41.57 & -6.04 & 2.46\\
\hline
\end{tabular}
\tablefoot{The parameters at the crossing period $P_{\mathrm{orb}}$ correspond to those the binary system is believed to have currently.\\
\tablefoottext{a}{Fraction of transferred mass lost from BH.}\\
\tablefoottext{b}{BH mass.}\\
\tablefoottext{c}{Donor mass.}\\
\tablefoottext{d}{Orbital period.}\\
\tablefoottext{e}{Fraction of hydrogen in core of the donor.}\\
\tablefoottext{f}{Age of donor.}\\
\tablefoottext{g}{BH spin.}\\
\tablefoottext{h}{Luminosity of donor.}\\
\tablefoottext{j}{Fraction of element H left in core.}\\
\tablefoottext{j}{Total weight for MT sequence from eq. \eqref{eq:likelihood}}\\
\tablefoottext{k}{Time spent around the observed orbital period.}
}
\end{sidewaystable*}
%\end{landscape}
%\restoregeometry

\begin{table*}
\centering
\caption{List of PPS property ranges at the RLO onset found from MT sequences with the analytic model and MESA respectively. Results from MESA has been divided into persistent and transient systems (see text).}
\begin{tabular}{cccc}
\hline
\hline
Parameter & Analytic model & \multicolumn{2}{c}{MESA} \\
\cline{3-4}
& & Persistent & Transient \\ 

\hline
$P_{\mathrm{RLO}}$ (days) & 0.70 - 2.80 & 0.80 - 1.70 & 0.80 - 1.70 \\
$M_{\mathrm{2, RLO}}$ ($M_{\mathrm{\odot}}$)        & 4.60 - 15.0  & 4.60 - 8.20 & 4.60 - 9.80 \\
$M_{\mathrm{BH, RLO}}$  ($M_{\mathrm{\odot}}$)  & 5.00 - 8.00  & 5.50 - 8.00 & 5.50 - 8.00 \\
$\beta$ & 0.00 - 1.00 & 0.00 - 0.90 & 0.90-1.00 \\
$\tau$ (Myr) & - & 33.53 - 88.52 & 36.09 - 87.90 \\
\hline
\label{tab:MTseq_range}
\end{tabular}
\end{table*}

In Table \ref{tab:set_winners} selected properties of both persistent and transient solutions to the RLO MT sequence of LMC X-3 are shown\footnote{The full table is available in the on-line version of this paper.}. Our grid of MT sequences is most successful in finding PPS when the RLO MT happens non-conservatively, with a donor star mass between $M_{\mathrm{2, RLO}} = 4.20-9.80\mathrm{M_{\odot}}$, a $M_{\mathrm{BH, RLO}} = 5.5 - 8.0 \mathrm{M_{\odot}}$, and an orbital period $P_{\mathrm{RLO}}$ = 0.8 - 1.7 days. In Table \ref{tab:MTseq_range} we outline the range of initial values at the moment of RLO onset for the donor mass, BH mass, mass fraction of lost material from the accretor and the current age of the system, as found from the analytic MT model, the persistent PPS, and the transient PPS. Overall the persistent PPS, relative to the transient PPS, can have a lower donor mass and several possible values of $\beta$. Transient solutions, on the other hand, require a slightly more massive donor and a high inefficient MT in order to be considered PPS.
\section{From ZAMS to RLO}
Before becoming a BH XRB, LMC X-3 began its evolution as a newly formed binary system, in a wide and potentially eccentric orbit. Given its high initial mass, the primary star quickly inflates as it evolves off the MS, eventually filling its Roche lobe and initiating MT on to the secondary. Due to the high mass ratio of the primary to the secondary star, the MT happens fast and is thermally and most likely dynamically unstable. In this dynamically unstable MT phase, also known as CE phase, the secondary star orbits within the tenuous envelope of the primary star, spirals inwards due to friction and heats the envelope. The CE phase results in either the merger of the two stars or in the ejection of the primary' envelope making the system a detached binary of a He core and a MS companion on a close and circular orbit.

%with the effect of shrinking the binary orbit and only a minor fraction of material is transferred to the secondary star. The much smaller orbit means the system is likely to go through a common envelope (CE) phase with the secondary star orbiting within the envelope of the primary star, potentially spiralling inwards, due to friction which also heat the envelope. The CE phase results in either the two stars merges or to the shredding of the primaries' envelope making the system a detached binary of a He core and a Main Sequence (MS) companion, where tidal interactions dissipates energy from the orbit that then shrinks and gets circularised.
Eventually the He core, left from the CE phase, goes into core-collapse and becomes a BH. During the SN explosion, asymmetries may alter the orbit further, potentially disrupting the system. If not, the binary system now comprises a BH and companion star, where tidal forces may change the orbit further. The companion star is largely unaffected during this process and continues its stellar evolution eventually reaching the onset of RLO. In this section, the range of PPS at the ZAMS stage that can lead to the appropriate characteristics at RLO onset is investigated.

\subsection{Population synthesis study with the Binary Stellar Evolution code}\label{sec:pop_synth}
During the formation of the BH binary, prior to the onset of RLO, many physical processes as described above take place. These processes are difficult to model, computationally expensive, and not yet fully understood from first principles. Even if all the physical processes could be modelled in detail, our estimate of the LMC X-3 progenitor properties at the onset of RLO can only provide a probable range, and is influenced by uncertainties in the observationally constrained parameters given in Table \ref{tab:LMC_X-3}. Therefore, when searching for PPS at the ZAMS, these uncertainties propagate backwards and increase further the range of PPS properties. Thus, a more approximate approach will be sufficient for the purpose of estimating the pre-RLO onset characteristics of LMC X-3 PPS. Such an estimate can be achieved with the Binary Stellar Evolution code \citep[BSE; ][]{hurley2002} which is a fast, parametric binary evolution code that allows simulating large sets of binary systems, accounting for all relevant physical processes such as stellar micro physics, CE, CO formation and SN dynamics.

We have modified BSE to include the suite of stellar wind prescriptions for massive stars described in \citet{belczynski2010}, the fitting formulae for the binding energy of the envelopes of stars derived by \citet{loveridge2011}, and the prescriptions "STARTRACK", "Delayed", and "Rapid" for CO formation in binaries, as described in \citet{fryer2012}. Each CO formation prescription favours a different distribution of CO masses as a function of ZAMS mass and corresponds to different formulations of the engines driving supernovae based on their metallicity, initial ZAMS mass, and pre-SN core mass. See \citet{fryer2012} for an elaborate description of the prescriptions.

As input to the population synthesis models we use the same set of $10^8$ binary systems which has been sampled as follows: 
The primary star $M_{\mathrm{1, init}}$ is sampled from a Kroupa initial stellar mass function \citep[IMF]{kroupa2001} with $\alpha$ = 2.3 and $M_{\mathrm{1, init}}=[20:40] \mathrm{M_{\odot}}$ and we assume in general that a Kroupa IMF can describe the distributions of stars of the LMC.
We draw a binary mass ratio $q_{\mathrm{init}}=$[0:0.5] from a uniform distribution, and combined with $M_{\mathrm{1, init}}$ gives the secondary stellar mass $M_{\mathrm{2, init}}$.
Thirdly, the orbital separation is sampled from a logarithmically uniform distribution in the range ]0:5] $(R_{\mathrm{\odot}})$. 
The fourth parameter sampled is the eccentricity and is drawn from the thermal eccentricity distribution \citep{jeans1919}. 
From these four parameters an initial orbital period is determined using Kepler's Third Law (eq. \ref{eq:orbital_period}).

A total of 10 different models were simulated with BSE, varying the CO mass prescription, the CE efficiency $\alpha_{\mathrm{CE}}$ = 0.0, 0.5, 1.0, and 2.0 and the distribution of kick velocities that BHs may receive during the SN. For the latter we considered a Maxwellian distribution with $\sigma_{\mathrm{v}}$  = 26.5, 100.0, and 265.0 km\,s$^{\mathrm{-1}}$, a uniform  distribution with an upper limit of 2000 km\,s$^{\mathrm{-1}}$, as well as the limiting case of no kicks. A Maxwellian distribution with 265.0 km\,s$^{\mathrm{-1}}$ is known to describe the natal kick distribution for single pulsars \citep{Hobbs:2005be} and here we explore whether it could also be qualitatively relevant for natal kicks in BH formation. By using different values for $\alpha_{\mathrm{CE}}$ and $\sigma_{\mathrm{v}}$ we avoid favouring one particular setup against another, hence recognizing that knowledge on both the CE phase and the natal kick distribution remain uncertain. Model 1 is chosen as our reference model (see Table \ref{tab:BSE_runs}).

\subsection{Population Synthesis Study Results}\label{sec:BSE}

\begin{table*}[!htb]
\centering
\caption{Setup of our population synthesis models using BSE (first 5 columns). The number of PPS found are shown in Col. 6. Column 7 shows the expected number of IMXB in the LMC today, see Sect. \ref{sec:IMXB}. CO prescription refers to those given in \citet{fryer2012}, $\alpha_{\mathrm{CE}}$ is the CE efficiency parameter. We have used two SN kick velocity distributions. For the Maxwellian distribution $\sigma_{\mathrm{v}}$ is the most probable kick velocity and for the uniform distribution it refers to the upper kick limit. $\sigma_{\mathrm{v}}=0$ means no natal kicks imparted onto the BH.}
\begin{tabular}{lcccccc}
\hline
\hline
Model 	& CO prescription	& $\alpha_{\mathrm{CE}}$  & Velocity distribution	 	& $\sigma_{\mathrm{v}}$ (km\,s$^{\mathrm{-1}}$) & N$_{\mathrm{PPS}}$ & N$_{\mathrm{IMXB}}$	\\
\hline
1 			& Rapid 				& 1.0 			& Maxwellian 					& 265.0						& 1322	 			& 0.72		\\
2 			& Rapid 				& 0.1 			& Maxwellian 					& 265.0						& 1 				& 0.21		\\
3 			& Rapid 				& 0.5 			& Maxwellian 					& 265.0						& 557 			    & 0.42		\\
4 			& Rapid 				& 2.0 			& Maxwellian 					& 265.0						& 5508				& 1.80		\\
5 			& Rapid 				& 1.0 			& Maxwellian 					& 0.0 						& 111 				& 0.43		\\
6 			& Rapid 				& 1.0 			& Maxwellian 					& 26.5 						& 161 				& 0.52		\\
7 			& Rapid 				& 1.0 			& Maxwellian 					& 100.0						& 980 				& 0.54		\\
8 			& STARTRACK 			& 1.0 			& Maxwellian 					& 265.0						& 29 				& 0.98		\\
9 			& Delayed				& 1.0 			& Maxwellian 					& 265.0						& 1222 				& 0.73		\\
%Note model 9 is run 15
10 		    & Rapid 				& 1.0 			& uniform 					    & 2000.0					& 245				& 	0.12	\\
%Note model 10 is run 23
\hline
\label{tab:BSE_runs}
\end{tabular}
\end{table*}

Each synthetic binary system was simulated from initial values at ZAMS, again using the metallicity Z = 0.006, forward to the companion star's RLO onset. Those systems that match to a MESA MT sequence model were stored with relevant information of the system at the ZAMS, the moment just prior to the SN explosion, the moment just after the explosion, and at the onset of RLO. The kick velocity $V_{\mathrm{kick}}$ and peculiar post-SN velocity $V_{\mathrm{pec, post SN}}$, following the explosion, were also recorded. We define a match between MESA MT sequence and a single BSE system if the system at the onset of RLO predicted by BSE are within 0.05 days of the orbital period, 0.1 $\mathrm{M_{\odot}}$ of the donor mass, and  0.25 $\mathrm{M_{\odot}}$ of the BH mass, of one of the MT sequences simulated with MESA. We only match BSE solutions with the 395 MT sequences identified as LMC X-3 PPS from modelling the X-ray MT phase, see Sect. \ref{sec:MTseq_results}. Of the 395 matching MT sequences, 121 are solutions to LMC X-3 as a transient source and the remaining 274 identified MT sequences are persistent solutions to LMC X-3.

The second rightmost column of Table \ref{tab:BSE_runs} shows the number of PPS for each population synthesis model. The reference Model 1 generated 1322 PPS out of a total of 10$^{\mathrm{8}}$ simulated binaries, suggesting at a first glance a relatively low formation rate, but see discussion in Sect. \ref{sec:pop_synth}. In model 2, 3, and 4 we vary the CE efficiency parameter, with $\alpha_{\mathrm{CE}}$ = 0.1, 0.5, 2.0 for each model respectively, and we find that the number of PPS increases with $\alpha_{\mathrm{CE}}$. In models 5, 6, and 7 we vary the distribution of natal BH kicks, with the most probable kick velocity being $\sigma_{\mathrm{v}}$  = 0.0, 26.5 and 100\,km\,s$^{\mathrm{-1}}$ for each model respectively, also finding an increasing number of PPS with increasing $\sigma_{\mathrm{v}}$. Model 8 adopts the "STARTRACK" prescription for the mass of CO formed, generating a small number of PPS, while in model 9 the "Rapid" prescription for CO formation was used, producing a number of PPS comparable to model 1.

The small number of PPS produced from models 2 and 8 are statistically insignificant samples of PPS to robustly estimate the range of LMC X-3 PPS at ZAMS. Based on the very low formation efficiency of these two models, we can only infer that it is plausible, but highly unlikely, that LMC X-3 is formed from the settings of these models. For LMC X-3 this suggests that the CE phase is necessary, and that the STARTRACK prescription is not a good estimate of CO masses. For model 4, which is the model that most efficiently forms PPS, it is hard to justify the high value of $\alpha_{\mathrm{CE}}$, but it also points towards the CE phase being important for generating LMC X-3 like systems. Since model 2 produced only 1 PPS we do not include it in the rest of our analysis.

\begin{figure*}%[!Htb]
\centering
\includegraphics[width=1.0\linewidth]{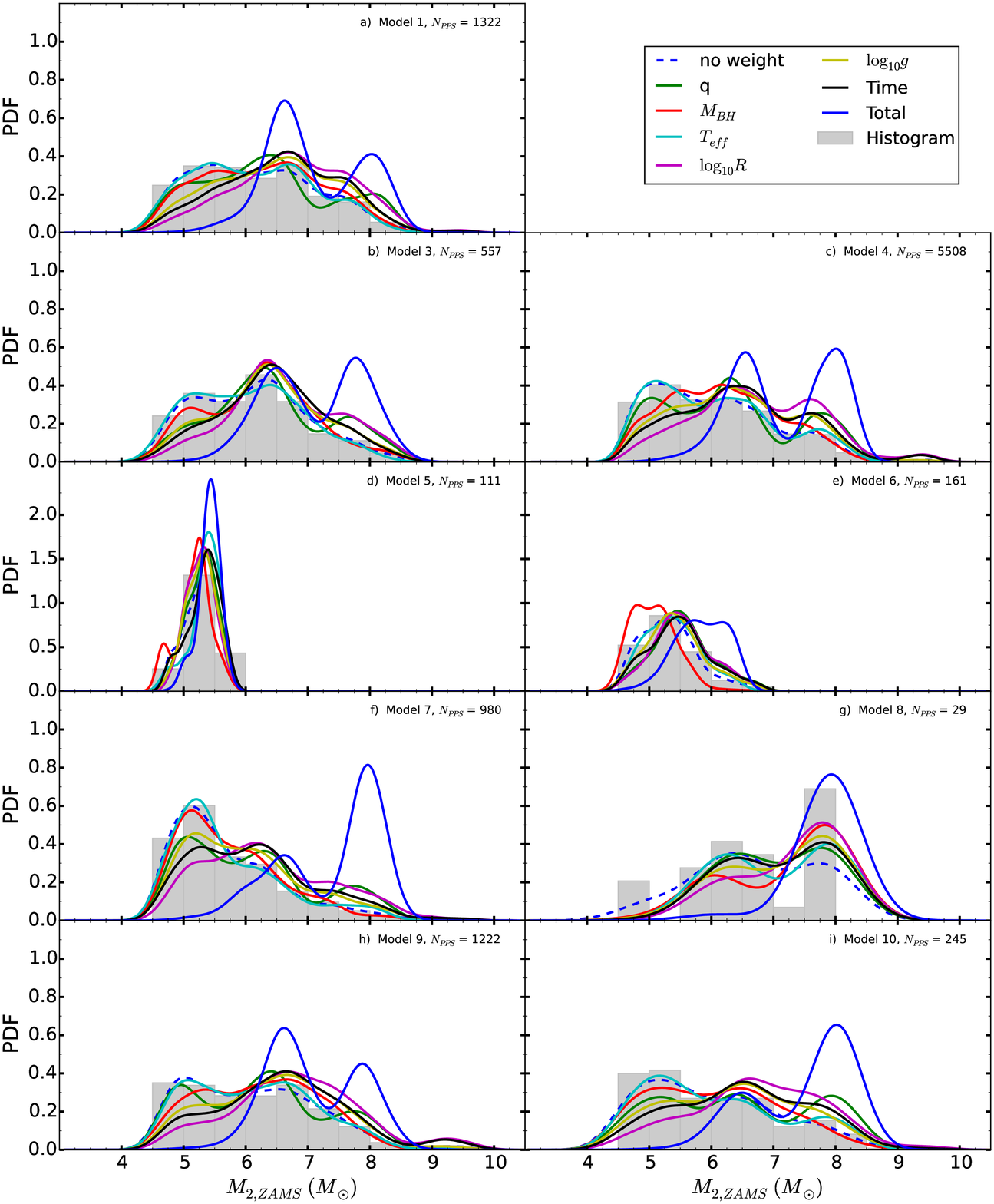}
\caption{The probability density function (PDF) of initial masses of the companion star $M_{\mathrm{2,ZAMS}}$ for the different synthesis population models in Table \ref{tab:BSE_runs}. The grey bar is a normalised histogram and the different colored lines are normalised Kernel Density Estimates for different weights found from eq. \eqref{eq:likelihood} and shown in the insert of this figure, for individual factors $j$ and for the total weight given as the product over all $j$ times the time spent around the orbital period, which is the black line.}
\label{fig:weight_demo}
\end{figure*}

Based on the number of PPS for each model, we perform a statistical analysis using weighted Kernel Density Estimators (KDE). A Gaussian kernel is used to produce a probability density estimate for each relevant property of PPS during the evolution of the LMC X-3. In Fig. \ref{fig:weight_demo} we demonstrate how the PDF of the companion stars' mass at ZAMS, $M_{\mathrm{2,ZMAS}}$, will look for each population synthesis model for different weights. The weights are found using a single factor in eq. \eqref{eq:likelihood}, i.e. only $\mathcal{I}_{\rm i,j}$ and a weight based on the time a MT sequence spends around the currently observed properties of LMC X-3. The latter is defined as the time a MT sequence spends within two standard deviations of all the observed constraints. Finally the total weight which is the product, i.e. left hand side, of eq. \eqref{eq:likelihood} $\mathcal{I}_{\rm i}$ multiplied by the time spent around the observed orbital period. The gray bar plot is a normalised histogram and the dashed line is a non-weighted KDE. The remaining lines are KDEs with different weights as described by the legend. It is shown that the unweighted KDE produces a continuous distribution that follows closely the discrete distribution of the histogram.
%Models 1, 3, 4, 7, 9, and 10 show comparable distributions whereas models 5, 6, and 8, likely due to their small numbers of PPS, produce much narrower distributions.
%The individual parameters, already analysed in the analytic MT model, i.e. the mass $M_{\mathrm{BH}}$ and $q_{\mathrm{eff}}$, follow more or less the unweighted distribution. %The two distributions using the weight of the radius $R$ and the donor mass $M_{\mathrm{2}}$ are almost impossible to discriminate from each other and this is independent on the population synthesis model. At first sight this might be surprising, but it should be expected from the fact that these two parameters in the RLO scenario are closely related, as it is the Roche lobe that controls the stellar radius and the amount of mass removed. 
The total weight eq. \eqref{eq:likelihood} over all $j$, relative to the other distributions, concentrates the most likely solution into two peaks except for model 5, 6, and 8. The most likely values based on the total weight also shows the two peaks to fall away from the peak of the unweighted distribution, the cause of which is the emergent distribution of using counts together with weights.
The distribution using the total weight finds a different maximum because the weights near the maximum with no weights are much smaller, by several orders of magnitude, than the weights at the maximum of the distribution with the total weight (see the color scale of Fig. \ref{fig:MTseq_no_Edd} which displays the values of the total weights). In the remaining analysis we use the total weight in all our distributions.

\begin{figure*}%[!Htb]
\centering
\includegraphics[width=1.0\linewidth]{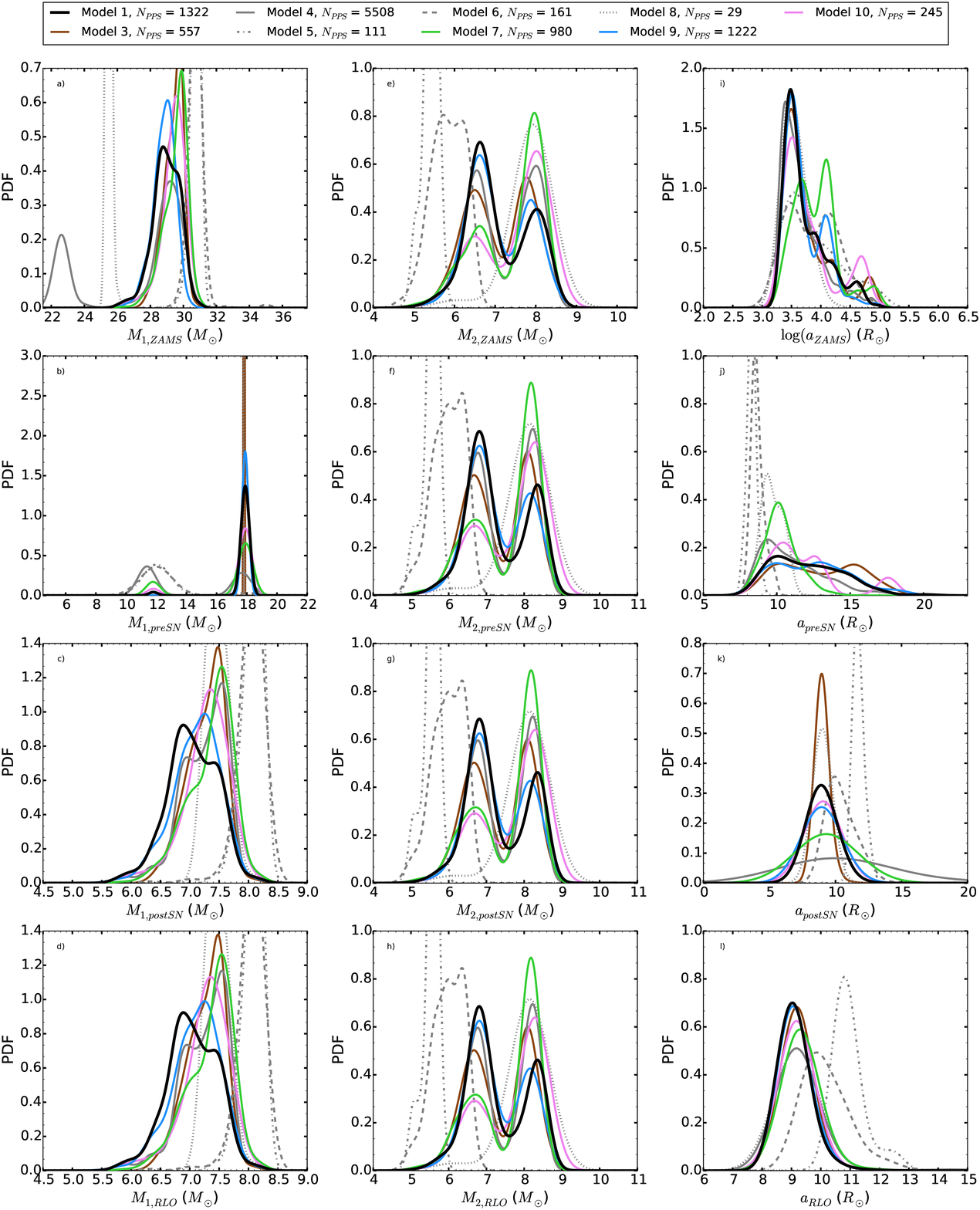}
\caption{The PDF of LMC X-3 progenitor's properties ($M_{\mathrm{1}}$, $M_{\mathrm{2}}$, and $a$) at different evolutionary stages, at ZAMS, pre SN, post SN, and onset of RLO, for the different synthesis population models in Table \ref{tab:BSE_runs}. The PDFs are weighted with the total weight.}
\label{fig:BSE}
\end{figure*}

%%%%%%%%%
Figure \ref{fig:BSE} shows the overall most likely evolution of LMC X-3, from its initial binary properties at ZAMS (top panel), to pre-SN (second row), post-SN (third row) and finally at RLO onset (bottom row), for the different population synthesis models, expressed as PDFs. The PDFs are produced as KDEs with the total weight applied, as defined earlier. Each color is a model from Table \ref{tab:BSE_runs} (see the figure legend). We have de-emphasized four models by presenting them in gray color.  Three of these (models 5, 6, and 8) because they have small or insufficient numbers of PPS to be statistical representable. Model 4 is also grayed out, but instead owing to its unphysical setting of $\alpha_{\mathrm{ce}} = 2$. The latter model was considered mostly in order to examine the sensitivity of the number of PPS to this parameter.

%%%%%%%%%
The remaining models 1, 3, 7, 9, and 10 have model parameters that are physically plausible and produce a sufficiently large number of PPS. These models estimate the values, at different evolutionary stages, of LMC X-3 progenitor's properties to be in the same range, along the x-axis of each panel. The peak of each distribution also seems to be at the same x-values, though, for places with a bimodal distribution, the individual model favors the peaks differently. %However, the peaks of the PDFs that each population synthesis model produces are distributed differently relative to each other, except for $a_{\mathrm{ZAMS}}$ and $M_{\mathrm{1,pre SN}}$ where all models suggest a peak around $\log(a_{\mathrm{ZAMS}})\sim$3.5\,$\mathrm{R_{\odot}}$ and $M_{\mathrm{1,pre SN}}\sim$17.8\,$\mathrm{M_{\odot}}$.

%%%%%%%%% M1
The initial value of $M_{\mathrm{1,ZAMS}}$ at the 95$\%$ percentile is in the range 25.4--31.0 $\mathrm{M_{\odot}}$ when considering all models except model 4. Model 4 has a range 22.4--30.1 $\mathrm{M_{\odot}}$. All models suggest that the primary mass just prior to the SN is close to $M_{\mathrm{1,pre SN}}\sim$17.8 $\mathrm{M_{\odot}}$, although models 4, 7, and 10 also suggest a minor peak around $\sim$12 $\mathrm{M_{\odot}}$. After the SN, the range of the CO masses is $M_{\mathrm{1, postSN}}$=6.4--8.2 $\mathrm{M_{\odot}}$ (95$\%$ percentile). %However, similarly to for the initial value of $M_{\mathrm{1,ZAMS}}$, model 3 favours a narrower mass range. 

%%%%%%%%% M2
The secondary star $M_{\mathrm{2}}$ experiences less dramatic changes, compared to the primary $M_{\mathrm{1}}$, evolving from a ZAMS to a BH. This is expected as $M_{\mathrm{2}}$ during this time is still on the MS and during the first MT episode the secondary mass star does not gain significant mass. There are two most probable values for the donor mass, one around 6.8\,$\mathrm{M_{\odot}}$ favoured by model 1,3,4, and 9. The second peak is around 8\,$\mathrm{M_{\odot}}$ and is favoured by models 7 and 10. The mass range spanned by all models is 5.0--8.4\,$\mathrm{M_{\odot}}$ (95$\%$ percentile).

%%%%%%%%% a
The orbital separation is initially very large, but as the primary star evolves away from the MS it fills its Roche lobe, and the binary gets into a CE phase, resulting in a significantly closer binary orbit. At the moment just prior to the SN, $a$ has decreased by a factor of $\sim$300, from several thousand solar radii to just around $\sim$10$\mathrm{R_{\odot}}$. The orbital separation following the SN, favours a value around $\sim$9$\mathrm{R_{\odot}}$.

\begin{figure}%[!Hb]
\includegraphics[width=1.0\linewidth]{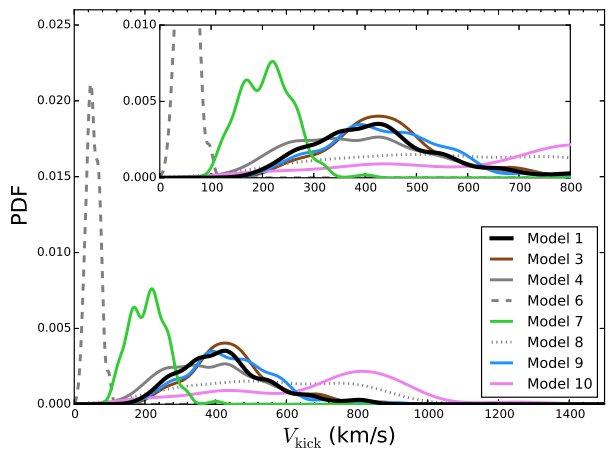}
\caption{The PDF of the kick imparted to LMC X-3 during the SN for the different synthesis population studies. The insert gives a more detailed view of the distribution in the x-axis range from 0 to 800~km/s and follows the same axis labels and plotted the lines the same legend.}
\label{fig:Vkick}
\end{figure}

\begin{figure}%[!Htb]
\includegraphics[width=1.0\linewidth]{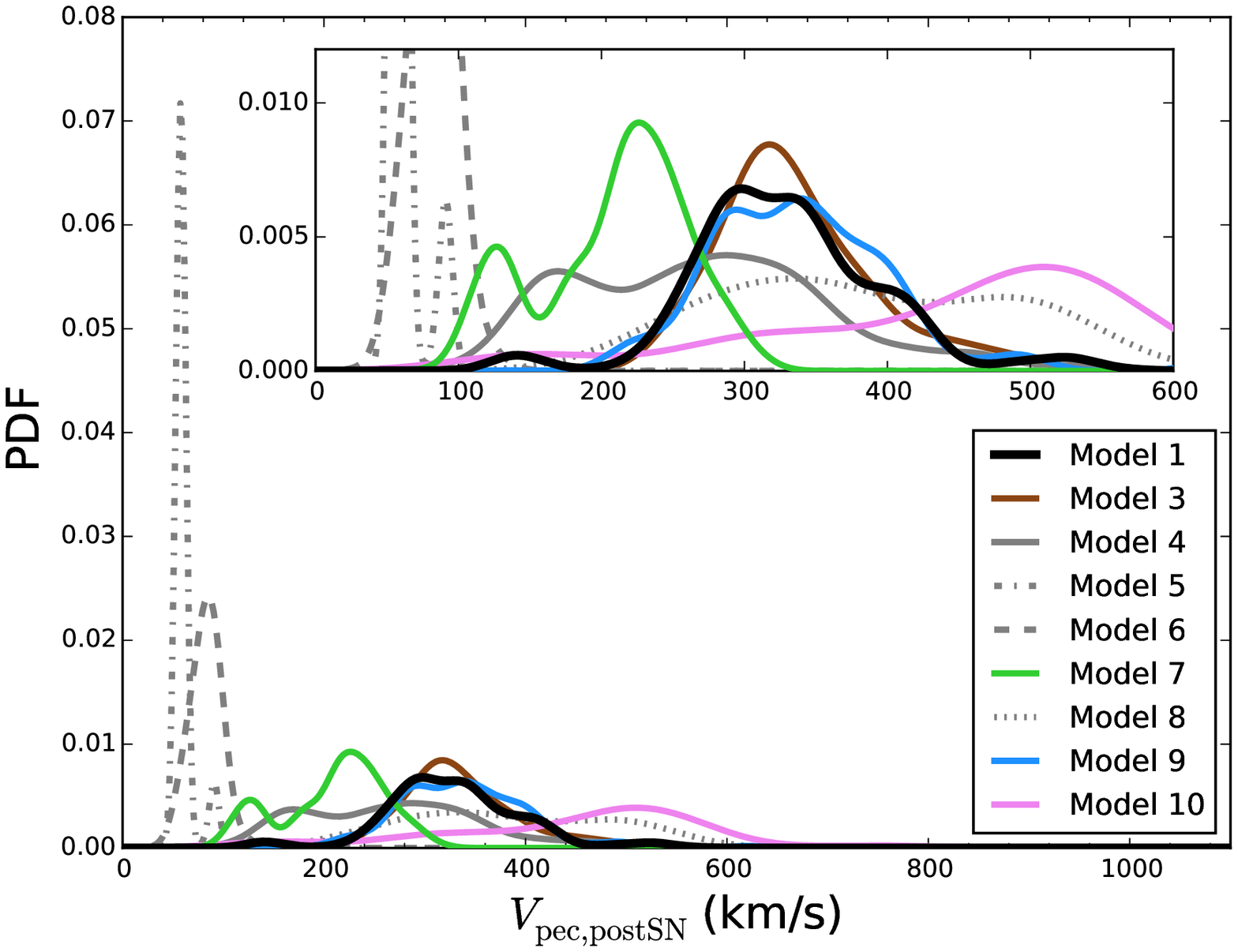}
\caption{The PDF of LMC X-3's center of mass velocity or  LMC X-3's peculiar post SN velocity, for the different population synthesis models. The insert gives a more detailed view of the distribution in the x-axis range from 0 to 600~km/s and follows the same axis labels and plotted the lines the same legend.}
\label{fig:Vsystem}
\end{figure}

The distribution of most likely kicks imparted during the BH formation is shown in Fig. \ref{fig:Vkick} and the post SN peculiar velocity shown in Fig. \ref{fig:Vsystem} which is due to both the kick imparted and the mass loss. 
Models 1, 3, 4, and 9 suggest the same distribution for the imparted kick and follow from the fact that these 4 models had the same input distribution. The range of these 4 models, is roughly between $V_{\mathrm{kick}}$=193--683\,km\,s$^{\mathrm{-1}}$ (95$\%$ percentile), with only models 5, 6, and 7 suggesting kicks less than 120\,km\,s$^{\mathrm{-1}}$, which is expected given the smaller most probable kick value for these models. Thus, the actual lower limit of $V_{\mathrm{kick}}$ may likely be closer to 120 \,km\,s$^{\mathrm{-1}}$. Notably, model 7, which uses $\sigma_{\mathrm{v}}=100$~km/s still shows a peak kick at $\sim 180$~km/s, which is supportive of the demand for an appreciable kick in the BH formation.
Model 10 applied an uniform kick distribution and finds PPS within the range $V_{\mathrm{kick}}$ = 212-912 \,km\,s$^{\mathrm{-1}}$ (95$\%$ percentile).
All in all the distribution of $V_{\mathrm{kick}}$ in the different population synthesis models indicates that a relative high kick velocity is likely to be imparted to the LMC X-3 progenitor system. We do find in our simulations that some models produce PPS with smaller or no kicks, but the associated probability is low relative to models favouring high kicks.
The consequence of the kick, assuming it does not disrupt the system, is a perturbation of the system's center of mass motion or the post SN peculiar velocity. In addition to the kick, material lost from the system via the primary also affects the post SN peculiar velocity $V_{\mathrm{pec, postSN}}$ \citep[][ see Fig. \ref{fig:Vsystem}]{kalogera1996, willems2005}, as this mass is ejected off center of the binary system center of mass.
Overall the $V_{\mathrm{pec, postSN}}$ in Fig. \ref{fig:Vsystem} is comparable in magnitude to the kick velocity shown in Fig. \ref{fig:Vkick}. For all models, except model 7, more than 90$\%$ of the PPS have $V_{\mathrm{kick}}$ > $V_{\mathrm{pec, postSN}}$. In model 6 and 7 only $\sim11\%$ and $\sim27\%$ have a larger kick velocity than $V_{\mathrm{pec, postSN}}$, and for models 5 the $V_{\mathrm{pec, postSN}}$ is always larger than the relating $V_{\mathrm{kick}}$. For model 10 with a uniform kick distribution $\sim99\%$ have $V_{\mathrm{kick}}$ > $V_{\mathrm{pec, postSN}}$.

%\newgeometry{left=1cm,right=1cm,top=0.5cm,bottom=0.5cm}

% \newgeometry{left=1cm,right=2cm,top=1.0cm,bottom=1.0cm}
% \begin{landscape}
\begin{sidewaystable*}%[hb]
\centering
\caption{View of the most likely range based on weighted 25$^{th}$ - 75$^{th}$ (5$^{th}$ - 95$^{th}$) percentiles, for each BSE model, one model per column. The total weight is applied.}
\begin{tabular}{lcccccccccc}
\hline
\hline
Model & 1 & 3 & 4 & 5 & 6 & 7 & 8 & 9 & 10 \\
\hline

$M_{1, ZAMS} (M_{\odot})$  & \makecell{28.5 - 29.5 \\ (27.7 - 30.1)} & \makecell{29.0 - 29.8 \\ (28.4 - 30.1)} & \makecell{26.5 - 29.4 \\ (22.4 - 30.1)} & \makecell{30.6 - 30.9 \\ (30.2 - 31.0)} & \makecell{30.6 - 30.9 \\ (30.0 - 31.0)} & \makecell{29.0 - 29.9 \\ (28.1 - 30.2)} & \makecell{25.4 - 25.5 \\ (25.4 - 25.7)} & \makecell{28.5 - 29.2 \\ (27.6 - 29.7)} & \makecell{29.0 - 29.9 \\ (28.0 - 30.2)} \\ 

$M_{2, ZAMS} (M_{\odot})$  & \makecell{6.5 - 7.7 \\ (5.8 - 8.2)} & \makecell{6.4 - 7.7 \\ (5.8 - 8.2)} & \makecell{6.4 - 7.9 \\ (5.8 - 8.3)} & \makecell{5.3 - 5.5 \\ (5.0 - 5.6)} & \makecell{5.4 - 6.1 \\ (5.0 - 6.4)} & \makecell{6.5 - 7.9 \\ (5.7 - 8.3)} & \makecell{7.8 - 8.0 \\ (6.0 - 8.0)} & \makecell{6.5 - 7.7 \\ (6.0 - 8.2)} & \makecell{6.5 - 8.1 \\ (6.0 - 8.2)} \\ 

$\log P_{ZAMS} (days)$  & \makecell{3.5 - 4.2 \\ (3.3 - 5.2)} & \makecell{3.5 - 4.2 \\ (3.3 - 5.5)} & \makecell{3.4 - 4.0 \\ (3.3 - 5.0)} & \makecell{3.5 - 4.6 \\ (3.3 - 5.6)} & \makecell{3.5 - 4.5 \\ (3.3 - 5.1)} & \makecell{3.7 - 4.4 \\ (3.3 - 5.5)} & \makecell{3.4 - 3.9 \\ (3.4 - 4.2)} & \makecell{3.5 - 4.3 \\ (3.3 - 4.7)} & \makecell{3.5 - 4.2 \\ (3.3 - 5.3)} \\ 

$e_{ZAMS}$  & \makecell{0.53 - 0.88 \\ (0.24 - 0.97)} & \makecell{0.46 - 0.87 \\ (0.23 - 0.98)} & \makecell{0.54 - 0.86 \\ (0.28 - 0.97)} & \makecell{0.64 - 0.95 \\ (0.38 - 0.99)} & \makecell{0.61 - 0.94 \\ (0.29 - 0.98)} & \makecell{0.72 - 0.92 \\ (0.36 - 0.99)} & \makecell{0.42 - 0.79 \\ (0.42 - 0.87)} & \makecell{0.56 - 0.89 \\ (0.28 - 0.95)} & \makecell{0.54 - 0.88 \\ (0.30 - 0.98)} \\ 

$a_{ZAMS} (\log R_{\odot})$ & \makecell{3.5 - 3.9 \\ (3.4 - 4.6)} & \makecell{3.4 - 3.9 \\ (3.4 - 4.8)} & \makecell{3.4 - 3.8 \\ (3.3 - 4.4)} & \makecell{3.5 - 4.2 \\ (3.4 - 4.9)} & \makecell{3.5 - 4.2 \\ (3.3 - 4.6)} & \makecell{3.6 - 4.1 \\ (3.4 - 4.8)} & \makecell{3.4 - 3.8 \\ (3.4 - 3.9)} & \makecell{3.5 - 4.0 \\ (3.4 - 4.3)} & \makecell{3.5 - 3.9 \\ (3.4 - 4.7)} \\ 

\hline
$M_{1, preSN} (M_{\odot})$ & \makecell{17.9 - 17.9 \\ (17.8 - 18.0)} & \makecell{17.8 - 17.8 \\ (17.8 - 17.9)} & \makecell{11.3 - 17.5 \\ (11.1 - 18.0)} & \makecell{12.0 - 12.1 \\ (11.9 - 17.9)} & \makecell{12.0 - 12.1 \\ (11.9 - 17.9)} & \makecell{17.9 - 17.9 \\ (11.7 - 18.0)} & \makecell{17.8 - 17.8 \\ (17.8 - 17.8)} & \makecell{17.9 - 17.9 \\ (17.8 - 17.9)} & \makecell{17.9 - 17.9 \\ (11.8 - 17.9)} \\ 

$M_{2, preSN} (M_{\odot})$ & \makecell{6.6 - 8.0 \\ (6.0 - 8.4)} & \makecell{6.6 - 8.0 \\ (6.0 - 8.4)} & \makecell{6.6 - 8.2 \\ (6.0 - 8.4)} & \makecell{5.4 - 5.6 \\ (5.0 - 5.8)} & \makecell{5.6 - 6.3 \\ (5.0 - 6.4)} & \makecell{6.6 - 8.2 \\ (5.8 - 8.3)} & \makecell{8.0 - 8.2 \\ (6.0 - 8.2)} & \makecell{6.6 - 8.0 \\ (6.0 - 8.4)} & \makecell{6.6 - 8.2 \\ (6.2 - 8.4)} \\ 

$P_{preSN} (days)$ & \makecell{0.7 - 1.2 \\ (0.6 - 1.6)} & \makecell{0.8 - 1.3 \\ (0.7 - 1.6)} & \makecell{0.7 - 1.1 \\ (0.6 - 1.5)} & \makecell{0.6 - 0.7 \\ (0.5 - 0.7)} & \makecell{0.7 - 0.7 \\ (0.6 - 0.8)} & \makecell{0.7 - 0.9 \\ (0.6 - 1.1)} & \makecell{0.6 - 0.7 \\ (0.5 - 0.8)} & \makecell{0.7 - 1.2 \\ (0.6 - 1.7)} & \makecell{0.8 - 1.0 \\ (0.6 - 1.6)} \\ 

$e_{preSN}$ & \makecell{0.07 - 0.16 \\ (0.03 - 0.23)} & \makecell{0.07 - 0.12 \\ (0.03 - 0.21)} & \makecell{0.00 - 0.01 \\ (0.00 - 0.23)} & \makecell{0.00 - 0.00 \\ (0.00 - 0.44)} & \makecell{0.00 - 0.00 \\ (0.00 - 0.32)} & \makecell{0.08 - 0.24 \\ (0.00 - 0.33)} & \makecell{0.02 - 0.03 \\ (0.02 - 0.03)} & \makecell{0.07 - 0.15 \\ (0.04 - 0.22)} & \makecell{0.07 - 0.21 \\ (0.00 - 0.28)} \\ 

$a_{preSN} (R_{\odot})$ & \makecell{10.0 - 13.8 \\ (8.9 - 16.4)} & \makecell{10.4 - 15.1 \\ (9.5 - 17.0)} & \makecell{9.2 - 12.2 \\ (8.5 - 15.2)} & \makecell{8.1 - 8.4 \\ (7.7 - 8.8)} & \makecell{8.4 - 8.9 \\ (8.1 - 9.7)} & \makecell{9.6 - 10.7 \\ (8.6 - 12.4)} & \makecell{9.2 - 10.3 \\ (8.0 - 10.4)} & \makecell{10.2 - 14.2 \\ (8.7 - 17.4)} & \makecell{10.0 - 12.6 \\ (9.0 - 17.0)} \\ 

\hline
$M_{1, postSN} (M_{\odot})$ & \makecell{6.8 - 7.4 \\ (6.4 - 7.7)} & \makecell{7.0 - 7.5 \\ (6.7 - 7.7)} & \makecell{7.0 - 7.5 \\ (6.6 - 7.7)} & \makecell{8.0 - 8.2 \\ (7.7 - 8.2)} & \makecell{7.9 - 8.2 \\ (7.6 - 8.2)} & \makecell{7.0 - 7.5 \\ (6.6 - 7.7)} & \makecell{7.3 - 7.5 \\ (7.3 - 7.7)} & \makecell{6.9 - 7.3 \\ (6.4 - 7.7)} & \makecell{7.0 - 7.5 \\ (6.5 - 7.7)} \\ 

$M_{2, postSN} (M_{\odot})$ & \makecell{6.6 - 8.0 \\ (6.0 - 8.4)} & \makecell{6.6 - 8.0 \\ (6.0 - 8.4)} & \makecell{6.6 - 8.2 \\ (6.0 - 8.4)} & \makecell{5.4 - 5.6 \\ (5.0 - 5.8)} & \makecell{5.6 - 6.3 \\ (5.0 - 6.4)} & \makecell{6.6 - 8.2 \\ (5.8 - 8.3)} & \makecell{8.0 - 8.2 \\ (6.0 - 8.2)} & \makecell{6.6 - 8.0 \\ (6.0 - 8.4)} & \makecell{6.6 - 8.2 \\ (6.2 - 8.4)} \\ 

$P_{postSN} (days)$ & \makecell{0.8 - 0.9 \\ (0.7 - 1.0)} & \makecell{0.8 - 0.9 \\ (0.7 - 1.0)} & \makecell{0.8 - 1.3 \\ (0.8 - 12.4)} & \makecell{1.2 - 1.3 \\ (1.2 - 1.4)} & \makecell{0.9 - 1.2 \\ (0.9 - 1.5)} & \makecell{0.8 - 0.9 \\ (0.8 - 1.1)} & \makecell{0.8 - 0.8 \\ (0.8 - 0.9)} & \makecell{0.8 - 0.9 \\ (0.7 - 1.0)} & \makecell{0.8 - 0.9 \\ (0.7 - 1.0)} \\ 

$e_{postSN}$ & \makecell{0.39 - 0.76 \\ (0.20 - 0.93)} & \makecell{0.45 - 0.78 \\ (0.16 - 0.95)} & \makecell{0.31 - 0.65 \\ (0.14 - 0.92)} & \makecell{0.28 - 0.29 \\ (0.28 - 0.32)} & \makecell{0.11 - 0.24 \\ (0.05 - 0.34)} & \makecell{0.18 - 0.42 \\ (0.09 - 0.57)} & \makecell{0.21 - 0.37 \\ (0.21 - 0.37)} & \makecell{0.43 - 0.75 \\ (0.19 - 0.94)} & \makecell{0.37 - 0.83 \\ (0.14 - 0.96)} \\ 

$a_{postSN} (R_{\odot})$ & \makecell{8.6 - 9.2 \\ (8.3 - 10.3)} & \makecell{8.6 - 9.2 \\ (8.2 - 9.9)} & \makecell{9.1 - 12.0 \\ (8.6 - 55.6)} & \makecell{11.5 - 11.9 \\ (11.3 - 12.4)} & \makecell{9.8 - 11.3 \\ (9.2 - 12.6)} & \makecell{8.9 - 9.6 \\ (8.4 - 10.3)} & \makecell{8.8 - 9.2 \\ (8.8 - 9.6)} & \makecell{8.7 - 9.2 \\ (8.3 - 10.0)} & \makecell{8.9 - 9.2 \\ (8.6 - 9.8)} \\ 

\hline
$V_{kick}$ (km/s) & \makecell{337.7 - 491.6 \\ (254.2 - 665.9)} & \makecell{378.2 - 483.3 \\ (258.0 - 683.7)} & \makecell{279.4 - 484.3 \\ (193.8 - 669.9)} & \makecell{0.0 - 0.0 \\ (0.0 - 0.0)} & \makecell{39.1 - 65.3 \\ (26.0 - 78.7)} & \makecell{171.0 - 251.9 \\ (123.6 - 303.3)} & \makecell{299.3 - 725.0 \\ (273.1 - 776.1)} & \makecell{364.0 - 519.1 \\ (257.8 - 624.5)} & \makecell{474.8 - 868.3 \\ (212.7 - 912.1)} \\ 

$V_{system}$(km/s) & \makecell{289.9 - 364.3 \\ (248.1 - 432.2)} & \makecell{304.0 - 368.5 \\ (257.6 - 457.5)} & \makecell{192.3 - 327.9 \\ (140.3 - 443.2)} & \makecell{56.9 - 59.4 \\ (54.2 - 91.3)} & \makecell{71.5 - 92.0 \\ (57.4 - 114.1)} & \makecell{180.1 - 250.3 \\ (112.6 - 287.0)} & \makecell{271.0 - 471.6 \\ (268.5 - 491.7)} & \makecell{295.8 - 381.4 \\ (245.1 - 432.6)} & \makecell{356.4 - 530.5 \\ (149.5 - 582.6)} \\ 

\hline
$M_{1, RLO} (M_{\odot})$ & \makecell{6.8 - 7.3 \\ (6.4 - 7.7)} & \makecell{7.0 - 7.5 \\ (6.7 - 7.7)} & \makecell{7.0 - 7.5 \\ (6.6 - 7.7)} & \makecell{8.0 - 8.2 \\ (7.7 - 8.2)} & \makecell{7.9 - 8.2 \\ (7.6 - 8.2)} & \makecell{7.0 - 7.5 \\ (6.6 - 7.7)} & \makecell{7.3 - 7.5 \\ (7.3 - 7.7)} & \makecell{6.9 - 7.3 \\ (6.4 - 7.7)} & \makecell{7.0 - 7.5 \\ (6.5 - 7.7)} \\ 

$M_{2, RLO} (M_{\odot})$ & \makecell{6.6 - 8.0 \\ (6.0 - 8.4)} & \makecell{6.6 - 8.0 \\ (6.0 - 8.4)} & \makecell{6.6 - 8.2 \\ (6.0 - 8.4)} & \makecell{5.4 - 5.6 \\ (5.0 - 5.8)} & \makecell{5.6 - 6.3 \\ (5.0 - 6.4)} & \makecell{6.6 - 8.2 \\ (5.8 - 8.3)} & \makecell{8.0 - 8.2 \\ (6.0 - 8.2)} & \makecell{6.6 - 8.0 \\ (6.0 - 8.4)} & \makecell{6.6 - 8.2 \\ (6.2 - 8.4)} \\ 

$P_{RLO} (days)$ & \makecell{0.8 - 0.9 \\ (0.8 - 1.1)} & \makecell{0.8 - 0.9 \\ (0.8 - 1.0)} & \makecell{0.8 - 0.9 \\ (0.8 - 1.3)} & \makecell{1.1 - 1.1 \\ (1.1 - 1.4)} & \makecell{0.9 - 1.1 \\ (0.8 - 1.4)} & \makecell{0.8 - 0.9 \\ (0.8 - 1.1)} & \makecell{0.8 - 0.8 \\ (0.8 - 0.9)} & \makecell{0.8 - 0.9 \\ (0.8 - 1.0)} & \makecell{0.8 - 0.9 \\ (0.8 - 1.0)} \\ 

$e_{RLO}$ & \makecell{0.00 - 0.00 \\ (0.00 - 0.00)} & \makecell{0.00 - 0.00 \\ (0.00 - 0.00)} & \makecell{0.00 - 0.00 \\ (0.00 - 0.00)} & \makecell{0.00 - 0.00 \\ (0.00 - 0.00)} & \makecell{0.00 - 0.00 \\ (0.00 - 0.00)} & \makecell{0.00 - 0.00 \\ (0.00 - 0.00)} & \makecell{0.00 - 0.00 \\ (0.00 - 0.00)} & \makecell{0.00 - 0.00 \\ (0.00 - 0.00)} & \makecell{0.00 - 0.00 \\ (0.00 - 0.00)} \\ 

$a_{RLO} (R_{\odot})$ & \makecell{8.8 - 9.4 \\ (8.5 - 10.3)} & \makecell{8.9 - 9.4 \\ (8.5 - 10.2)} & \makecell{8.9 - 9.5 \\ (8.5 - 11.9)} & \makecell{10.6 - 11.1 \\ (10.4 - 12.5)} & \makecell{9.7 - 10.8 \\ (9.2 - 12.3)} & \makecell{9.0 - 9.7 \\ (8.5 - 10.5)} & \makecell{8.9 - 9.4 \\ (8.9 - 9.8)} & \makecell{8.8 - 9.4 \\ (8.5 - 10.2)} & \makecell{8.9 - 9.4 \\ (8.7 - 10.0)} \\ 

\hline
\end{tabular}
\label{tab:BSEresults}
\end{sidewaystable*}
% \end{landscape}
% \restoregeometry%get the old one pack

\subsection{Example of full evolution from ZAMS to White Dwarf formation}
\begin{figure*}%[!Htb]
\centering
\includegraphics[width=1.0\linewidth]{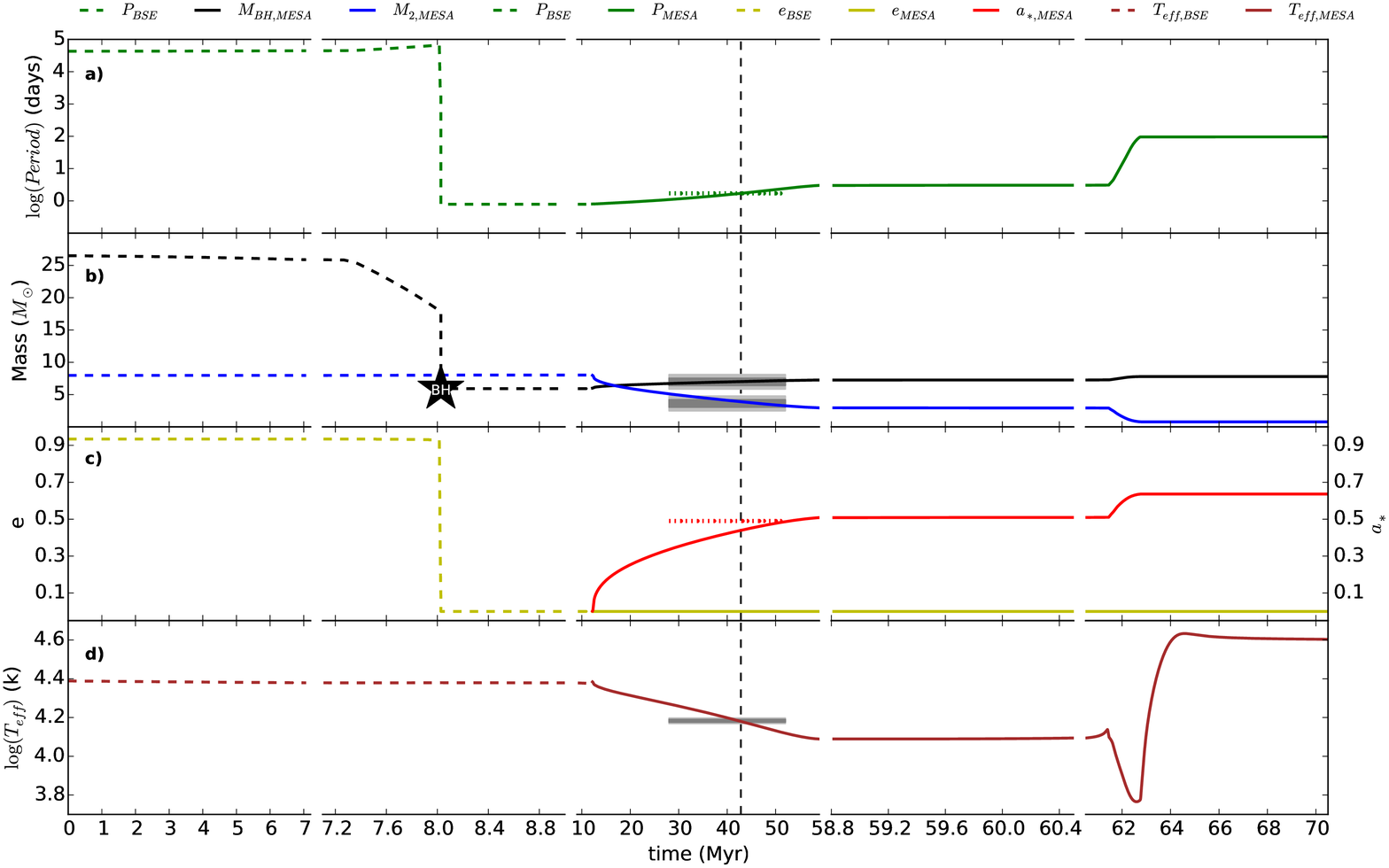}
\caption{Illustration of the complete evolution, of a typical LMC X-3 PPS from ZAMS to a double CO (BH+WD) binary. Panel a) shows the evolution of the orbital period in days on a logarithmic scale. Panel b) displays the evolution of the masses in the binary system; the primary star $M_{\mathrm{1}}$ as a star and as a BH (black line), and the companion star $M_{\mathrm{2}}$ (blue line). The black star label indicates the time of BH formation. Panel c) shows the evolution of the system's eccentricity $e$ in yellow, while in red we show the BH spin parameters $a_{\ast}$. Panel d) shows the effective temperature $T_{\mathrm{eff}}$ of the donor star in a logarithmic scale. Data coming from BSE simulation are shown as dashed lines, while data from MESA as solid lines. From table \ref{tab:LMC_X-3} we have added; in panel a) the observed orbital period as the green dotted line, in panel b) the 1$\sigma$ and 2$\sigma$ error of $M_{\mathrm{2}}$ and $M_{\mathrm{BH}}$ centered on the observed values, in panel c) the red dotted line is the upper 2$\sigma$ limit on the BH spin parameter, and in panel d) the 1-$\sigma$ and 2$\sigma$ error of $T_{\mathrm{eff}}$. In general, there is a fine agreement between BSE and MESA at the moment of RLO, except for the age of the system at RLO onset. MESA calculations find that RLO onset occurs at time = 8.8 Myr, whereas BSE finds it to begin at time = 12.2 Myr. The reason for the discrepancy is the differences in the micro physics, between BSE and MESA, and in particular in the stellar structure calculations. Here, for illustrative purposes, the MESA calculations are offset to the right by 2.4 Myr in order to generate a smooth transition in the figure. The star marks the formation of the BH at $\sim$8 Myr and the vertical dashed line marks the time $\sim$42 Myr at which the system crosses the observed orbital period.}
\label{fig:single_PPS}
\end{figure*}

In Fig. \ref{fig:single_PPS} we illustrate the complete evolution of a typical LMC X-3 PPS, from ZAMS and until the companion star becomes a WD. The system chosen is the MESA MT sequence with relative high likelihood for the X-ray binary phase and one of its BSE matches for the evolution prior to the X-ray phase, hence representative of our results. 
The evolution begins with a primary star of mass $M_{1,ZAMS}$ = 26.5 $\mathrm{M_{\odot}}$ and its companion star with mass  $M_2$=8\,$\mathrm{M_{\odot}}$ in a wide and highly eccentric orbit. The primary star evolves quickly and fills its Roche lobe, leading to a CE phase which shrinks and circularises the orbit. During the CE phase the donor star spirals inwards within the primary envelope. Due to friction the envelope is heated and expelled leaving behind a detached binary of a naked He core and the donor star. The remaining He core of the primary eventually experiences a SN explosion at $\sim8$ Myr and collapses into a BH. During the SN the BH receives a natal kick $V_{kick}$ = 497\,km\,s$^{\mathrm{-1}}$. 4.2 Myr later at time = 12.2 Myr, the companion star fills its Roche lobe and mass starts flowing from the donor star onto the BH. The system is now an observable X-ray source until a time of 58.8 Myr where the MT briefly stops for a few Myr. At $\sim$62 Myr MT is initiated again for a short period of time ($\sim$0.5 Myr). The brief pause is the end of the donor's MS evolution. During this pause the donor star contracts and restructures its interior and the system is developing from a case A MT into a case B MT. Following the pause the donor star is in a shell H-burning phase with a He-core. Due to expansion of the orbit and thus the companion star's radius during the shell-burning, as well as the intense MT, the donor's $T_{\mathrm{eff}}$ drops to a minimum value. Eventually, when the envelope of the companion star is almost completely removed, the companion star contracts well within its Roche lobe and the binary detaches. At this point the MT stops and the effective temperature of the companion star increases fast indicating that what is left is a degenerate low mass helium core on its way to become a helium WD. The final binary consists of a 12.1 $\mathrm{M_{\odot}}$ BH with spin $a_{\ast}=0.63$, and a 0.6 $\mathrm{M_{\odot}}$ He WD in a 96 days orbit.

\subsection{The expected number of IMXB within LMC today}\label{sec:IMXB}

In our reference model (Model 1 in Table \ref{tab:BSE_runs}), which is the most efficient in producing LMC X-3 PPS (excluding the unphysical models), we find 1322 PPS out of $10^8$ models binary systems. We find fewer PPS for the model with the "Delayed" CO formation prescription (Model 9), and even fewer with the "STARTRACK" prescription (Model 8) or small natal kicks. Here we address the question of whether we can expect to observe a LMC X-3 like system today within the LMC, knowing the formation efficiency of LMC X-3 of the different BSE models.

To do so, we define an IMXB as any binary system with a BH and a companion star of mass $3 \leq M_2/M_{\odot} < 10$ going through RLO MT. To estimate the number of IMXBs produced in the LMC within some period of time T, we make a population synthesis model for a new sample of binary systems. The new sample follows the same distributions as described in Sect. \ref{sec:pop_synth}, but with the range of the Kroupa IMF set to $M_{\mathrm{1,ZAMS}}$=[10,100] $\mathrm{M_{\odot}}$ and $q_{\mathrm{init}}$ = ]0,1]. This ensures, that the new sample is sufficiently broad to not only produce a subset of binary systems relevant for a subset of IMXB but can produce all IMXB within our definition. In the end, we must correct our identified number of IMXB $N_{BSE,IMXB}$ with the expected fraction of stars within the sample range $f_{IMF}[10;100]$. Making the sample sufficiently large further ensures a statistical significant number of IMXBs is produced. The sample size is $N_{\mathrm{BSE,sim}}$=2$\times$10$^{\mathrm{7}}$ binary systems. The number of IMXB produced within the LMC during the time period T is then given as
\begin{equation}
N_{IMXB, LMC}(T) = \frac{N_{BSE,IMXB}}{N_{BSE,sim}}f_{IMF}[10;100]N_{bin}(T),
\end{equation}
where $N_{bin}(T)$ is the number of binary systems produced in the LMC within a time period T. From our definition of IMXB, the population synthesis model yields a maximum IMXB age to be $\sim$300Myr, hence we set T = 300 Myr. $N_{bin}$ is then estimated by assuming a mean star-formation rate in the LMC over the past 300 Myr of $\sim$0.2\,$\mathrm{M_{\odot}\,yr^{-1}}$ \citep{harris2009} and a mean stellar mass of 0.518 $M_{\odot}$ based on a binary fraction of 70\%, a Kroupa IMF, and a uniform mass ratio distribution within binary systems. We get $N_{bin} = 0.7 \times 0.5 \times 0.2\mathrm{M_{\odot}\,yr^{-1}} \times 300 \mathrm{Myr} / 0.518 \mathrm{M_{\odot}}$ = $4\times10^7$.
The IMF correction $f_{IMF}[10;100]$ = 0.0014 is found by integrating the normalised IMF in the mass range [10;100] $\mathrm{M_{\odot}}$. The number of produced IMXB within LMC in the past 300 Myr is then $N_{IMXB, LMC} = 4.45$ systems.

To get the expected number of observable IMXBs today within the LMC, $N_{IMXB, LMC, Obs}$, we multiply with the mean life time of IMXBs, $\tilde{\tau}_{IMXB}$, i.e. the time that they are X-ray bright as IMBXs,  over the relevant period T:
\begin{equation}
N_{IMXB, LMC, Obs} = N_{IMXB, LMC}\frac{\tilde{\tau}_{IMXB}}{T}. 
\end{equation}
From the BSE simulation used to estimate the number of IMXBs formed, we also derive the mean lifetime of IMXBs to be $\tilde{\tau}_{IMXB}$ = 50.2\,Myr for model 1. Hence, the expected number of observable IMXBs within the LMC for model 1 is 0.72 system. This is consistent with the fact that we only observe one IMXB system in LMC today, which is LMC X-3. Column 7 of Table \ref{tab:BSE_runs} yields the result for all BSE models.
\section{Discussion}

\subsection{The natal spin of the BH}

In the analytic MT model introduced in Sect. \ref{sec:analytic_model}, we made the explicit assumption that the natal spin of the BH is zero. This is motivated from the recent work by \citet{2015ApJ...800...17F}. They suggested that BH progenitor stars of Galactic BH LMXBs loose most of their angular momentum before the onset of the CE, due to wind mass-loss and envelope expansion during the giant phase, and showed that the currently observed BH spin in Galactic LMXBs can be explained solely due to accretion after the BH formation. However, direct observational or theoretical evidence that the natal spin of BHs in Galactic LMXBs is negligible, is not yet available.

\begin{table}
\centering
\caption{Same as Table \ref{tab:MTseq_range}, but for different values of  natal BH spin}
\begin{tabular}{cccc}
\hline
\hline
Parameter & $a_{*,natal}=0$ & $a_{*,natal}=0.1$ & $a_{*,natal}=0.2$ \\
\hline
$P_{\mathrm{RLO}}$ (days) & 0.8 - 1.7 & 0.8 - 1.7 & 0.8 - 1.7 \\
$M_{\mathrm{2, RLO}}$ ($M_{\mathrm{\odot}}$)    & 4.6 - 9.8  & 4.6 - 9.8 & 4.6 - 9.8 \\
$M_{\mathrm{BH, RLO}}$  ($M_{\mathrm{\odot}}$)  & 5.5 - 8.0  & 5.5 - 8.0 & 5.5 - 8.0 \\
$\beta$ & 0.0 - 1.0 & 0.0 - 1.0 & 0.0 - 1.0 \\
$\tau$ (Myr) & 33.53 - 87.90 & 33.55 - 87.9 & 36.09 - 87.9 \\
\hline
\label{tab:MTseq_range_natal_spin}
\end{tabular}
\end{table}

In the context of our analysis, the assumption of zero natal spin is in fact a conservative one. This is because we only require the mass that the BH accretes from the onset of the RLO phase until its current state, does not spin up the BH to a value of $a_*$ that is higher than two sigma above the observed value. If a non-zero natal BH spin is assumed, then the maximum amount that the BH in LMC X-3 may have accreted after its formation is further constrained to a smaller value. This has as a direct effect that the minimum allowed natal mass of the BH goes up, but should in principle also narrow down the allowed range of all the binary properties of the system at the onset of RLO. Table \ref{tab:MTseq_range_natal_spin} shows how these ranges change for a choice of natal BH spin of $a_{*,natal}=0.1$ and $a_{*,natal}=0.2$, relative to our standard assumption of zero natal spin. Given, however, that it remains an open question how exactly BH form and how the formation process may affect their natal spin, we decided to adopt the most conservative, wider range of allowed RLO properties, that corresponds to a natal BH spin of $a_{*,natal}=0$.
As Table \ref{tab:MTseq_range_natal_spin} shows, althorugh individual MT sequences may not qualify anymore as PPS, there are close to no changes in the overall estimated ranges. Only the minimum present age of the system increases as the natal spin increases and is barely noticeable for $a_{\ast}=0.1$. The minimum age increases, as the more massive donors in a small orbit will begin their RLO early enough to spin up the BH above the allowed spin limit.
Assuming no natal spin we find a total of 395 MT sequences as PPS for LMC X-3, and for a natal spin of 0.1 and 0.2 respectively 371 and 336 PPS.
As a limiting case, we may consider the BH spin to be entirely natal, hence accretion onto the BH would not be possible. This would be equivalent to our case of MT sequences with $\beta = 1$.

\subsection{The X-ray flux of LMC X-3}
\begin{figure}[!Htb]
\centering
\includegraphics[width=1.0\linewidth]{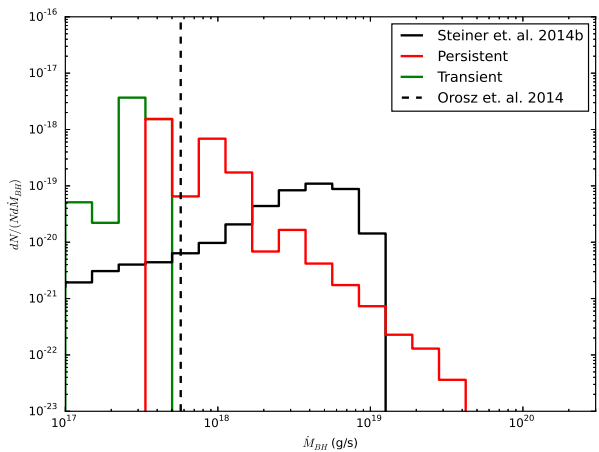}
\caption{Distribution of observationally determined instantaneous accretion rates onto the BH taken at different time (black), and distribution of long-term average BH accretion rates inferred from MT sequence, calculated with MESA, that successfully reproduce the observed characteristics of LMC X-3. The green curve corresponds to MT sequences that predict a transient XRB, while the red curve corresponds to persistent ones. The vertical black dashed line shows the critical accretion rate for the thermal disk instability model, given by eq. \eqref{eq:accept_criteria_5}, for the observationally inferred properties given in Table \ref{tab:LMC_X-3}.}
\label{fig:Mdot}
\end{figure}

In Fig. \ref{fig:Mdot} we compare the accretion rates onto the BH as inferred from the \textit{winning} MESA MT sequence models from Sect. 3.4 to the observed values from \citet{steiner2014x-ray_lag}. The black solid line is the distribution of the observed instantaneous accretion rates, while the green and the red lines correspond to the long-term average accretion rate estimates from transient and persistent MESA MT sequences, correspondingly. The black dashed vertical line is the critical accretion rate for the development of the disk thermal instability, given from eq. \eqref{eq:accept_criteria_5}, using the observed values of LMC X-3 from Table \ref{tab:LMC_X-3}.

Overall, it is shown that the observed accretion rates covers a larger range where as the MT sequence models distributes themselves with most likely accretion rates near the critical accretion rate given by eq. \eqref{eq:accept_criteria_5} for the observed properties of LMC X-3 from Table \ref{tab:LMC_X-3}. This is similar to what was shown in Fig. \ref{fig:MTseq}, where the three examples of MT sequences were close to the critical accretion rate. We should stress here that a direct comparison between the observations and the model predictions is not possible. This is because the observed instantaneous accretion rates cover a baseline of $\sim10$ yrs, while our modelled MT sequences yield an average value over timescales of few 10$^{\mathrm{5}}$ to several 10$^{\mathrm{6}}$ yrs and do not include any potential short term fluctuations. Keeping this in mind, the general agreement between our predictions and the data at hand are notable and suggest a positive affirmation of our model. The fact that our MT calculations predict MT rates in the transition region between transient and persistent sources provides a natural explanation for the atypical X-ray variability of LMC X-3.

\subsection{Mechanism of Compact Object Formation}
Although we see quantitative differences between the three different CO prescriptions, "Rapid", "Delayed", and "STARTRACK" as a SN mechanism to have produced LMC X-3 we cannot on this basis make a global ruling between the three prescriptions. It is however evident that "STARTRACK" is not able to reproduce a statistical significant set of PPS for LMC X-3. Both the "Rapid" and "Delayed" prescriptions produce comparable results. A relevant question to ask is how the three prescriptions perform in producing a population of IMXB as we did in Sect. \ref{sec:IMXB}. The result of repeating the population synthesis study (Sect. \ref{sec:IMXB}) for IMXB in LMC using the "STARTRACK" prescription yields that 0.98 IMXB is currently observable within the LMC. The "Delayed" prescription yields a total of 0.73 systems is currently observable. In conclusion all three prescriptions give a consistent number of IMXB in LMC relative to observations. The same can be said for the other individual models in table \ref{tab:BSE_runs} with varying $\alpha_{\mathrm{ce}}$ and $\sigma_{\mathrm{v}}$.

\subsection{Natal kicks and the orbit of LMC X-3 within the LMC}
An interesting perspective to be drawn from our synthesis population study with BSE is the increase in the PPS number, as we increase $\sigma_{\mathrm{v}}$ of the assumed Maxwellian asymmetric kick distribution from 0 \,s$^{\mathrm{-1}}$ to 265km\,s$^{\mathrm{-1}}$. Few PPSs are found when no kick or just a small kick is imparted. 
%In both cases though, $\sigma_{\mathrm{v}}=0\,\rm km\,s^{-1}$ or $\sigma_{\mathrm{v}}=26.5\,\rm km\,s^{-1}$, {\red our inferred probability of observing any IMXB in LMC falls below XX}. 
By gradually increasing the kick to be imparted until it equals the NS kick distribution, we increase the number of PPSs. Most noticeable is the jump when going from $\sigma_{\mathrm{v}}$=26.5 \,s$^{\mathrm{-1}}$ to $\sigma_{\mathrm{v}}$=100\,km\,s$^{\mathrm{-1}}$, where the number of PPSs increases by a factor $\sim$10. Although we do not have direct evidence, there is strong indication that LMC X-3 likely has formed with a kick being imparted to it. Given the lack of precise enough peculiar motion measurements of LMC X-3, which would allow us to put direct constraints on the natal kick that the BH received, our estimates are depended on the assumed model for the compact object formation. This is because the latter dictates the relation between the pre-SN mass of the BH progenitor and its natal BH mass, and hence the amount of mass lost during SN. A high mass-loss during SN tends to result in wide post-SN orbits, for which in turn a large kick in the appropriate direction is required in order to keep the tight.

The estimated post SN peculiar velocity or systemic velocity of the binary's center of mass is also relatively high. Even for cases with zero or small kicks the systemic velocity is likely above 50\,km\,s$^{\mathrm{-1}}$. This is due to the high mass loss from the primary during the SN (see Figure 5). This relative large post SN system velocity opens a question which only a kinematic study of the LMC X-3 within LMC could answer. Given the high systemic velocity the system potentially received, it should affect the system's orbit within the parent galaxy. So, how is this orbit looking today? If one had sufficiently accurate measurements of the systems 3-dimensional motion and knew the potential of LMC, it could be tested whether LMC X-3 is in a normal orbit, most likely suggesting a low kick and low systemic velocity. If instead LMC X-3 was found to have a peculiar orbit, it would suggest that the system would have received a high kick and high systemic velocity. Finally, if the LMC X-3 was in an extreme orbit or perhaps even escaped the LMC it would indicate that the LMC X-3 received a very high kick. In this respect it might be possible to constrain the lower or upper natal kick imparted.

For LMC X-3 only the radial velocity is measured with accuracy. If one look at observations of radial velocities, which are somewhat survey dependent, they suggest a radial velocity of $V_{\mathrm{rad}}$ = 300-308 km s$^{\rm -1}$; see Table 1 in \citet{orosz2014}. Using the model of \citet{vandermarel2002} updated with the numbers of \citet{vanderMarel:2014bi}, one can get the expected radial velocity of a system in the LMC. This model gives for LMC X-3 an expected radial velocity of 244 km s$^{\rm -1}$ which compared to actual observations yields a difference of $\sim$55km s$^{\rm -1}$ suggesting the LMC X-3 center of mass rest frame is moving faster than its neighbouring stellar population even when accounting for a velocity dispersion of $\sim$20km s$^{\rm -1}$ \citep{vanderMarel:2014bi}.
The proper motion of LMC X-3 is $\mu_{\alpha}cos(\delta)$ = -2.7 mas yr$^{\mathrm{-1}}$ $\pm$ 24.4 and $\mu_{delta}$ = 12.5 $\pm$ 24.4 mas/yr$^{\mathrm{-1}}$ \citep{Smart:2014ff}, which if translated into a 3d velocity yield uncertainties so high it cannot be used to constrain the LMC X-3 past evolution.

\subsection{LMC X-3: The young sibling of GRS 1915+105?}\label{sec:1915}
A comparison of our MT sequences from MESA to the Galactic LMXB GRS1915+105, which has the largest measured spin of all LMXBs $a_{\ast}>0.98$;\citep[][]{mcclintock2006grs1915105}, provides some interesting insights.
Overall GRS1915+105's current properties are P = 33.85 days \citep{Steeghs2013}, $M_{\mathrm{BH}} = 12.4 \pm 2.0 \mathrm{M_{\odot}}$ \citep{reid2014grs1915105}, $M_{\mathrm{2}} = 0.52 \pm 0.41 \mathrm{M_{\odot}}$ \citep{privateMcClintock}, donor's spectral type K0III-K3II and $T_{\mathrm{eff}} = 4766.5 \pm 666.5$\,K \citep{Harlaftis:2004bo,Greiner:2001ce, Gray:2008wb, Anonymous:dO47EiUo}.

\begin{figure*}[!Htb]
\centering
\includegraphics[width=1.0\linewidth]{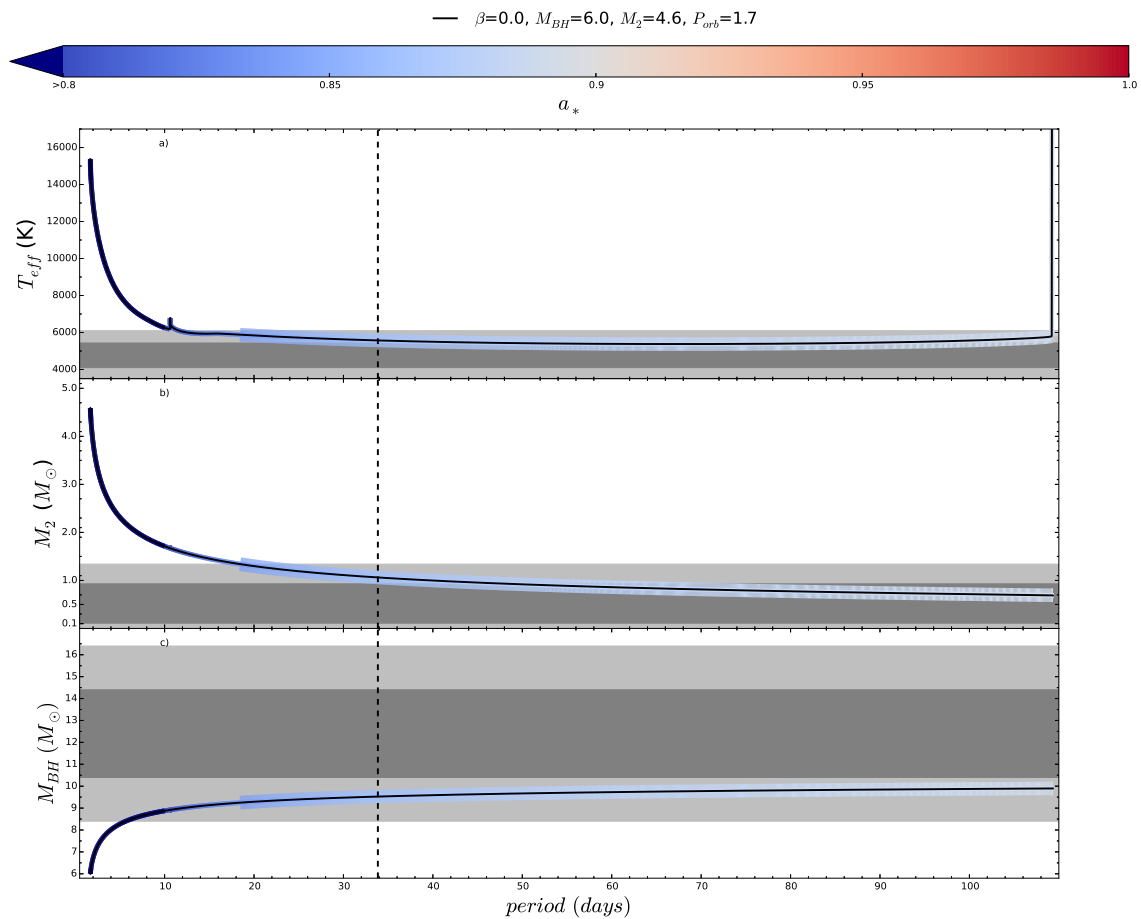}
\caption{A winning MT sequences simulated with MESA from RLO onset (see legend for values) and onwards until the donor becomes a WD. In the three panels, the observed properties of the Galactic LMXB GRS 1915+105 are shown, indicated by the gray areas. The light and dark gray areas are centered on the observed value of GRS 1915+105 and each gray contrast span 1$\sigma$ on each side, except for $M_{\mathrm{2}}$ where the lower 2$\sigma$ limit is truncated at 0. The dashed vertical line is the observed orbital period of 33.85 days of GRS1915+105. The color demonstrates the BH spin parameter $a_{\ast}$ which for GRS 1915+105 is $a_{\ast}>0.98$, and it shows the maximum spin to be below 0.9. The thickening of the color illuminating the MT seq in each panel indicates the part of the evolutionary track for which the model properties are within 2$\sigma$ of the observed properties of GRS~1915+105 (excluding the BH spin). The sudden spike seen in the top panel is the formation of a WD.}
\label{fig:GRS1915}
\end{figure*}

For our comparison we reran LMC X-3's winning MT sequences with $\beta$ = 0, allowing the largest amount of mass transferred as possible, and compared them to the parameters of GRS~1915+105 at its current orbital period. For the masses $M_{\mathrm{BH}}$, $M_{\mathrm{2}}$, and $T_{\mathrm{eff}}$ we accept models that are within $2\sigma$ at the observed orbital period. For the BH spin parameter we make no constraint. 
We find 11 MT sequence models that match GRS1915+105 parameters and with BH spin in the range 0.76 to 0.86. Figure \ref{fig:GRS1915} show the comparison of one MT sequence model (see legend for settings at RLO onset). In the three panels are shown $T_{\mathrm{eff}}$, $M_{\mathrm{2}}$, $M_{\mathrm{BH}}$ as a function of orbital period. The color bar shows the BH spin parameter, as the BH is accreting material. Once the color thickens the properties of the MT sequence model are within 2$\sigma$ of GRS~1915+105's currently observed properties. The evolution of the system is initially similar to that described in section 3.3. The abrupt increase in $T_{\mathrm{eff}}$ at the end of the evolution is the formation of a WD.
The similarity between the evolution of LMC X-3, from our simulations, and GRS1915+105 indicate that these two systems potentially have formed from similar progenitor binary systems. Finally, although winning MT sequences for LMC X-3 with $\beta=0$ can reach BH spins $a_\ast > 0.8$, it is clear that the combination of donor and BH masses at the onset of RLO do not provide a large enough mass reservoir to spin up the BH to an extreme spin. As shown in \citet{2015ApJ...800...17F}, the range of donor masses at RLO onset that are required in order to spin up the BH in GRS~1915+105 to $a_{\ast}>0.98$ is slightly higher than our inferred range for LMC X-3. Relaxing our assumption for negligible natal spin for the case of GRS~1915+105 would yield a much larger overlap between the PPSs for the two systems. This on the other hand would indicate a different compact object formation process for the BHs in the two systems.

\section{Conclusion}
In a two part study, we have reconstructed the evolution of LMC X-3 from its current X-ray bright MT phase, back to when the system was at the ZAMS. In the first part we formulated an analytic MT point-mass model for the XRB phase that includes binary component mass and orbital angular momentum changes, as well as changes in the spin parameter of the BH. The analytic MT point-mass model was used to initially limit the possible parameter space of LMC X-3's PPS at the moment of RLO onset, for which we then did detailed numerical simulations with MESA. We computed a regular space grid of 3319 MT sequences of which 395 are solutions of the RLO onset settings of LMC X-3, with 121 MT sequences predicting a transient XRB and 274 MT sequences predicting a persistent one. From our detailed MT sequences we found that LMC X-3's current MT rate is very close to the critical MT rate predicted by the thermal instability disk model, which defines the boundary between a persistent and transient source. This could potentially explain LMC X-3's high X-ray variability along with the fact that the system has always been bright since its discovery.

We then used the range of solutions at RLO onset found from MESA to match to a large set of binary population synthesis simulations done with BSE. We considered 10 different models with BSE, and for each model we estimated the most likely range for PPSs of LMC X-3 at the ZAMS, just prior to the SN, just after the SN, and at the onset of RLO onset. In addition we recorded the natal kick imparted to the BH during the SN and the systemic velocity after the kick. Overall we find, at 95~$\%$ confidence across all models, that LMC X-3 began as a ZAMS binary with a primary of mass $M_{\mathrm{1,ZAMS}}$ = 22.4-31.0 $\mathrm{M_{\odot}}$, a secondary at $M_{\mathrm{2,ZAMS}}$ = 5.0-8.2 $\mathrm{M_{\odot}}$, an orbital period of $P_{\mathrm{ZAMS}} = 2\,000-400\,000$ days, and an eccentricity $e_{\mathrm{ZAMS}}$ $\geq$ 0.23. Just prior to the SN, the primary had a mass $M_{\mathrm{1,preSN}}$ = 11.1-18.0 $\mathrm{M_{\odot}}$, with the secondary star largely unaffected, though it might have accreted some of the mass lost from the primary. From ZAMS to just prior to the SN, the orbit decreased to periods between 0.6-1.7 days, and $e_{\mathrm{preSN}} \leq$ 0.44. We do find systems with no natal kick imparted, but these are few, and it is much more likely that the SN explosion imparted a large kick of 120\,km\,s$^{\mathrm{-1}}$ or more. Following the SN, the system has a BH of $M_{\mathrm{postSN}}$ = 6.4-8.2 $\mathrm{M_{\odot}}$ in an eccentric orbit. From the time just after the SN and until the RLO onset, the orbit is circularised and it has an orbital period $P_{\mathrm{RLO}}$ = 0.8-1.4 days.

Finally, we discussed the potential for LMC X-3's evolution to be similar to the much older Galactic LMXB GRS1915+105. We demonstrated how LMC X-3 will likely not build up an equally extreme spin, because the LMC X-3 system overall has too little mass to accumulate such an extreme spin through accretion. Despite this, it is likely that GRS1915+105 was also an IMXB when its RLO began, and followed a similar evolution starting with a slightly higher mass companion star.

In summary, we have successfully traced the evolution of LMC X-3 from a ZAMS binary system until its current state. Studying LMC X-3 and constraining its characteristics, especially during the CO formation, is of great importance for the understanding of how BH IMXB form and evolve. From an observational point of view, we also offer a likely explanation for the high X-ray variability associated with LMC X-3, by estimating the current, long-term average MT rate close to the boundary between transient vs. persistent behaviour. Our prediction that the system likely received a large natal kick and has a correspondingly large post SN peculiar velocity could be tested with sufficient kinematic data and a proper motion study.

\begin{acknowledgement}
This work was supported by the Swiss National Science Foundation (project number 200020-160119). M.S. is grateful for support from the Swiss National Science Foundation and to the StarPlan Center, University of Copenhagen where much of this work was performed. T.S. acknowledges support from the Ambizione Fellowship of the Swiss National Science Foundation (grant PZ00P2_148123). J.F.S. has been supported by NASA Einstein Fellowship grant PF5-160144. The computations were performed at University of Geneva on the Baobab cluster. All figures were made with the free Python module Matplotlib\citep{matplotlib}.
\end{acknowledgement}
%\clearpage
\bibliographystyle{apalike}  %{apalike}, plain,gji
\bibliography{mads_library,tassos_library}
%\clearpage
\end{document}